\documentclass{article}

\usepackage{microtype}
\usepackage{graphicx}
\usepackage{subfigure}
\usepackage{booktabs} 

\usepackage[accepted]{template_double_column/icml2021}

\icmltitlerunning{Submission and Formatting Instructions for ICML 2021}
\usepackage{cite}
\usepackage{amsmath,bm,amssymb,amsfonts,mathtools}
\usepackage{latexsym}
\usepackage[switch]{lineno}

\usepackage[utf8]{inputenc}
\usepackage{array,multirow,graphicx}
\usepackage{adjustbox}
\usepackage[table,xcdraw]{xcolor}
\usepackage[separate-uncertainty = true, multi-part-units = single]{siunitx}
\usepackage{tabularx}
\usepackage{soul, xcolor}
\usepackage[normalem]{ulem}

\usepackage{url}
\usepackage{xcolor}
\definecolor{newcolor}{rgb}{.8,.349,.1}
\usepackage[colorlinks]{hyperref}
\hypersetup{
colorlinks=true,
  }
\definecolor{DarkGreen}{rgb}{0.2,0.5,0.2} 
\makeatletter
\definecolor{airforceblue}{rgb}{0.36, 0.54, 0.66}
\definecolor{babyblue}{rgb}{0.54, 0.81, 0.94}
\definecolor{bluegray}{rgb}{0.4, 0.6, 0.8}
\AtBeginDocument{\def\@citecolor{bluegray}}
\AtBeginDocument{\def\@linkcolor{bluegray}}

\usepackage[colorinlistoftodos,prependcaption,textsize=small]{todonotes}

\usepackage{graphicx}
\usepackage{import}

\def\BibTeX{{\rm B\kern-.05em{\sc i\kern-.025em b}\kern-.08em
    T\kern-.1667em\lower.7ex\hbox{E}\kern-.125emX}}

\usepackage{multirow}
\usepackage{tabularx}
\usepackage{hhline}
\usepackage{array}
\newcolumntype{P}[1]{>{\centering\arraybackslash}p{#1}}
\newcolumntype{L}[1]{>{\raggedright\let\newline\\\arraybackslash\hspace{0pt}}m{#1}}
\newcolumntype{C}[1]{>{\centering\let\newline\\\arraybackslash\hspace{0pt}}m{#1}}
\newcolumntype{R}[1]{>{\raggedleft\let\newline\\\arraybackslash\hspace{0pt}}m{#1}}
\usepackage{booktabs}

\usepackage{paralist}

\usepackage[normalem]{ulem}
\usepackage{xfrac}

\RequirePackage[normalem]{ulem} 
\usepackage{soul}
\usepackage{ulem}
\usepackage{censor}
\newlength\nextcharwidth
\makeatletter
\renewcommand\@cenword[1]{%
  \setlength{\nextcharwidth}{\widthof{#1}}%
  \censorrule{\nextcharwidth}%
  \kern -\nextcharwidth%
  #1}
\makeatother

\providecommand{\DIFadd}[1]{{{{\protect{#1}}}}} 
\providecommand{\DIFdel}[1]{{}}    

\usepackage{enumitem}
\usepackage{graphicx}
\usepackage{amsmath}
\usepackage{amssymb}
\usepackage{booktabs}
\usepackage{chngcntr}  
\usepackage{listings}

\usepackage{threeparttable}  
\graphicspath{{figures/}}
\DeclareGraphicsExtensions{.pdf,.eps}

\usepackage[acronym,nomain]{glossaries}
\makeglossaries
 
\newacronym{ace}{ACE-Net}{A-line Coordinates Encoding Network}
\newacronym{sbert}{Slot-BERT}{}
\newacronym{bert}{BERT}{}
\newacronym{vae}{VAE}{Variational Auto-encoder}
\newacronym{sa}{SA}{Slot Attention}
\newacronym{pca}{PCA}{Principal Component Analysis }
\newacronym{nlp}{NLP}{Natural Language Processing}
\newacronym{vit}{ViT}{Vision Transformer}
\newacronym{tst}{TST}{Temporal Slot Transformer}
\newacronym{top}{TOP}{Transient Object Presence}
\newacronym{hd}{HD}{Hausdorff distance}

\begin{document}

\twocolumn[
\icmltitle{Slot-BERT: Self-supervised Object Discovery in Surgical Video}






\icmlsetsymbol{equal}{*}
\begin{icmlauthorlist}
\icmlauthor{Guiqiu Liao}{pcaso}
\icmlauthor{Matjaz Jogan}{pcaso}
\icmlauthor{Marcel Hussing}{cis}
\icmlauthor{Kenta Nakahashi}{tgh}
\icmlauthor{Kazuhiro Yasufuku}{tgh}
\icmlauthor{Amin Madani}{sai}
\icmlauthor{Eric Eaton}{cis}
\icmlauthor{Daniel A. Hashimoto}{pcaso,cis}
 
\end{icmlauthorlist}

\icmlaffiliation{pcaso}{Penn Computer Assisted Surgery and Outcomes Laboratory,
 Department of Surgery, University of Pennsylvania, Philadelphia, PA, USA.}
\icmlaffiliation{cis}{Department of Computer and Information Science, University of Pennsylvania, Philadelphia, PA, USA.}
\icmlaffiliation{tgh}{Division of Thoracic Surgery, Toronto General Hospital, University Health Network, Toronto, Ontario, Canada.}
\icmlaffiliation{sai}{Surgical Artificial Intelligence Research Academy, University Health Network, Toronto, ON, Canada.}

\icmlcorrespondingauthor{Guiqiu Liao}{guiqiu.liao@pennmedicine.upenn.edu}

\icmlkeywords{Machine Learning, ICML}

\vskip 0.3in
]



\printAffiliationsAndNotice{}  

\begin{abstract}
Object-centric slot attention is a powerful framework for unsupervised learning of structured and explainable representations that can support reasoning about objects and actions, including in surgical video. While conventional object-centric methods for videos leverage recurrent processing to achieve efficiency, they often struggle with maintaining long-range temporal coherence required for long videos in surgical applications. On the other hand, fully parallel processing of entire videos enhances temporal consistency but introduces significant computational overhead, making it impractical for implementation on hardware in medical facilities.
We present Slot-BERT, a bidirectional long-range model that learns object-centric representations in a latent space while ensuring robust temporal coherence. Slot-BERT scales object discovery seamlessly to long videos of unconstrained lengths. A novel slot contrastive loss further reduces redundancy and improves the representation disentanglement by enhancing slot orthogonality.
We evaluate Slot-BERT on real-world surgical video datasets from abdominal, cholecystectomy, and thoracic procedures. Our method surpasses state-of-the-art object-centric approaches under unsupervised training achieving superior performance across diverse domains. We also demonstrate efficient zero-shot domain adaptation to data from diverse surgical specialties and databases. Code and data shared at: \href{https://github.com/PCASOlab/slot-BERT}{https://github.com/PCASOlab/slot-BERT}.
\end{abstract}

\section{Introduction}
\label{sec:intro}

Research suggests that humans learn to perceive the world through object-specific grouping and association  \citep{kahneman1992reviewing,tenenbaum2011grow}. This ability constructs representations akin to object files that bind and track features of an object over time thereby enabling more efficient use by downstream cognitive tasks. Inspired by this cognitive process, self-supervised object-centric learning aims to learn explainable and adaptable representations from unlabeled datasets~\citep{burgess2019monet,greff2019iodine,locatello2020object,greff2020binding}.

While object-centric learning comes in different forms, one particularly effective  {approach} incorporates the inductive bias of grouping or binding low-level, unstructured perceptual activations into a set of vectors known as slots. Each slot encapsulates a higher-level compositional entity, such as an object. This grouping or binding process emerges from architectural priors combined with self-supervised learning techniques, including attention and auto-encoding through specialized encoder-decoder architectures. Substantial progress has been made in this field for both image \citep{seitzer2022bridging,fan2024adaptive,mansouri2023object,jiang2023object,wu2023slotdiffusion} and video \citep{weis2020unmasking,kipf2021conditional,aydemir2023self,zadaianchuk2024object,wu2022slotformer,biza2023invariant,lee2024guided,qian2023semantics,bao2022discovering,bao2023object,singh2022simple,singh2024parallelized} processing.

In object-centric image processing, features like color and semantic embeddings from pre-trained models are used to learn  pixel-to-object assignments.
Methods proposed for object-centric processing of video add additional cues about the temporal consistency of objects, including optical flow \citep{kipf2021conditional} and between-frame similarities \citep{zadaianchuk2024object} or depth maps \citep{elsayed2022savi++}. While beneficial, these additional cues can lead to increased computational complexity and a higher risk of error accumulation. For instance, optical flow often fails with static or deformable objects and large inter-frame displacements, while depth maps may be not available for certain video domains and unreliable in low-light or low-contrast settings. Feature similarity can address certain challenges but may result in homogeneity when training on smaller datasets as the similarity map is agnostic to the semantics of these features.

These issues become especially pronounced in {learning disentangled representation for scene decomposition in} surgical videos, where temporal dynamics can be highly complex. Instruments and tissues in surgical videos often move at different speeds, and the visibility of specific objects can fluctuate significantly over the course of a procedure. Moreover, many current object-centric learning approaches for videos fail to fully exploit the sequential structure {of} longer { episodes. In surgical videos, a typical episode depicting a surgical action or a surgical task will last from a few seconds to a minute or longer.}  \citep{kipf2021conditional, singh2022simple, elsayed2022savi++} struggle with learning from {sequences of this length}.  
Employing slot-attention mechanisms to process entire video sequences {in parallel}, as proposed in \citep{singh2024parallelized}, could help overcome some challenges but may struggle with scalability for longer videos in real-world datasets due to computational limitations. 

{These limitations underscore the need for an architecture that is computationally tractable for longer video sequences and supports self-supervised training without auxiliary cues which are not inherently robust (e.g., optical flow or between-frame patch similarity). An ideal model would operate in the latent space that supports scalable temporal reasoning well beyond the short effective context of auto-regressive RNNs.}

We propose an architecture that is both easy to train and excels in long-range bidirectional temporal reasoning to efficiently handle longer video sequences {---} \gls{sbert}, an self-supervised object-centric slot attention model that performs temporal reasoning over slots using a bidirectional transformer encoder. The model treats slots learned from video frames as foundational visual concepts, similar to how text embeddings represent words. By adapting the BERT transformer encoder from natural language processing to process encoded video slots (akin to a sequence of word embeddings), \gls{sbert} learns to reconstruct video feature maps, providing implicit supervision for object discovery through masked auto-encoding of slots. This approach implements a temporal bidirectional self-attention mechanism, enabling effective video representation learning. Furthermore, by learning to reconstruct masked slots, \gls{sbert} can predict future slots, and we can use this functionality for future initialization of slots that improves the accuracy when applied on longer sequences.
We also propose a video slot contrastive loss that increases the independence between slots within a video by maximizing orthogonality in the latent vector space. This regularization leads to a more distinct representation of concepts and increases the precision of segmentation maps.

To summarize, our contributions are as follows: 
 
\begin{itemize}
     \item {We introduce \gls{sbert}, a novel object-centric self-supervised representation learning model based on bidirectional temporal reasoning across video frames.}
      \item  {We introduce slot-contrastive loss, specifically designed for slot attention, to improve orthogonality between slots}. 
       \item  {Our model is computational efficient and easy to train on longer videos, and can run on affordable hardware.}
      \item { We demonstrate superior temporal coherence, bidirectional reasoning, and zero-shot generalization compared to state-of-the-art methods across four surgical video datasets from three different domains}.

   \end{itemize}
\section{Related Work}
\label{sec_related_work}
%
\subsection{Self-supervised object-centric learning} 
\label{unsupervised_object_centric}
Learning of object-centric representations can demonstrably improve the sample efficiency and generalization of vision and dynamics models in compositional domains. In order to discover objects from images or videos without supervision, the learning architecture must bind the distinct features that belong to particular objects and create representations with activation patterns that are distinct for object instances. Various approaches have been proposed to achieve this goal. Early works focused on variational auto-encoders (VAEs) \citep{kingma2013auto} and disentangled representation learning \citep{mathieu2019disentangling,eastwood2018framework,kim2018disentangling,higgins2017beta}. Other efforts employed iterative attention mechanisms, such as Capsule Networks \citep{sabour2017dynamic,hinton2018matrix,tsai2020capsules}, contrastive learning approaches \citep{kipf2019contrastive,henaff2022object,xu2022groupvit}, or complex-valued auto-encoders that implicitly encode temporal correlation \citep{lowe2022complex}. Beyond learning generalized representations, reconstructive and generative object-centric models also show the ability to segment objects in scenes \citep{lin2020space,van2020investigating,greff2019iodine,engelcke2019genesis,burgess2019monet}.

\gls{sa}~\citep{locatello2020object,kori2024identifiable}, a recently proposed iterative attention mechanism, proved effective in grouping latent representations of objects into a number of slots, and can scale to real-world scenes by reconstructing pre-trained features from foundation models like DINO \citep{caron2021emerging} or MAE \citep{he2022masked}. \gls{sa} can use multiple modalities to discover objects, such as optical flow \citep{kipf2021conditional}, depth maps \citep{elsayed2022savi++}, temporal feature similarity \citep{zadaianchuk2024object}, or 3D point clouds \citep{ibrahim2023sat3d}. Furthermore, \gls{sa} can embed object entities in an identifiable manner \citep{kori2024identifiable}, with promising results demonstrated on synthetic data. It has also been applied to improve Vision-Language Modeling~\citep{xuslotVLM} and controllable synthetic image generation \citep{jiang2023object,wu2023slotdiffusion}. 

In video processing, association of slots across time requires initialization and \gls{sa} on video typically uses RNN-like architectures \citep{kipf2021conditional,zadaianchuk2024object,singh2022simple,elsayed2022savi++} that support iterative initialization of slots. These approaches however suffer from instability when training on longer sequences and in scenes with a limited temporal coherence. A parallelized approach, such as \citep{singh2024parallelized}, processes the entire video sequence as a batch, which comes at the cost of computational efficiency and makes it less suitable for real-world video applications.

Our \gls{sa} architecture builds on the RNN framework by incorporating a temporal fusion transformer to enhance long-range bidirectional reasoning{, which is inspired by BERT for language modeling \mbox{\citep{kenton2019bert}}. We treat the slot representation as a language token and adopt the same pre-training strategy that is used in training of language models to the training of video \gls{sa} models.   }This novel design improves temporal coherence and maintains scalability for longer video sequences while addressing the limitations of prior methods.

%

\subsection{Masked Information Encoding}

Masked visual modeling has been proposed to learn effective visual representations, starting with training of denoising auto-encoders where masks were treated as noise~\citep{vincent2010stacked}. Transformer architectures ~\citep{vaswani2017attention}  brought significant progress in \gls{nlp} \citep{kenton2019bert, radford2018improving} and vision \citep{dosovitskiy2021image, arnab2021vivit} thanks to learned dot product attention for long-range encoding of co-occurrences.  
Building upon the success of masked self-attention in GPT \citep{radford2018improving}, iGPT learned to predict sequences of pixels~\citep{chen2020generative} or masked tokens as in the ViT model~\citep{dosovitskiy2021image}

The success of these vision transformers inspired a number of other transformer-based architectures for masked visual modeling \citep{bao2022beit, dong2021peco, wei2022masked}. BEiT \citep{bao2022beit} and BEVT \citep{wang2022bevt} followed the language model BERT \citep{kenton2019bert} learning visual representations from images and videos by predicting discrete tokens \citep{ramesh2021zero}. Image MAE \citep{he2022masked} and Video MAE \citep{tong2022videomae} used asymmetric encoder-decoder architectures for masked image modeling based on plain ViT backbones. MaskFeat \citep{wei2022masked} proposed reconstructing HOG features of masked tokens to perform self-supervised pre-training in videos.  

Masked-out features or tokens have also been applied to weakly supervised settings, where class labels were provided while the objective was to produce attention maps for detection and segmentation \citep{hou2018self, lee2019ficklenet}. 

Inspired by BERT, our method treats slots encoded from images as analogous to word embeddings, incorporating a transformer to process video slots. The training objective is to recover image features when slots are partially masked out. Following this strategy, the transformer module can perform reasoning across time instead of merely duplicating information from individual slots.


%
\subsection{Unsupervised Object Detection and Segmentation in Video}
\label{subsec_semantic_seg}

Unsupervised object detection and segmentation in videos often rely on motion cues to identify objects and regions of interest. Motion-based approaches have been extensively applied to video segmentation \citep{fragkiadaki2015learning,Kossen2020Structured, ponimatkin2023simple,liu2021emergence,karazija2022unsupervised,choudhury2022guess}. For example, Fragkiadaki et al. \citep{fragkiadaki2015learning} ranked segment proposals by combining optical flow with static boundary information. Similarly, Tokmakov et al. \citep{tokmakov2017learning} utilized optical flow to capture motion cues; however, their method struggles to segment static objects due to the lack of detailed spatial information.  

To address these challenges, MATNet \citep{zhou2020motion} incorporated motion information to enhance spatiotemporal object representation. While this fusion of static and motion data improved performance, it encountered difficulties with complex moving backgrounds and was highly dependent on the accuracy of optical flow maps. In the domain of surgical videos, Sestini et al. \citep{sestini2023fun} employed a teacher-student network to refine optical flow-based segmentation, improving accuracy in these specialized scenarios.  

Beyond motion-based methods, Croitoru et al. \citep{croitoru2019unsupervised} proposed using \gls{pca} to isolate foreground objects in a fully unsupervised manner. For cases where the segmentation of the first frame is available, tracking-based methods \citep{uziel2023vit,cheng2023tracking} can effectively propagate segmentation across video frames. To effectively address the issue of moving background and uncertainty of object presence in surgical videos, an approach using feature ranking and knowledge distillation is proposed \citep{liao2024disentangling}, however this approach needs weak supervision using video-level class labels.

Our work focuses on object-centric learning that produces explainable representations and slot-based segmentation masks at the same time. While we leverage motion cues for object discovery through inductive biases, our method can learn moving and static objects using a unified framework. By relying solely on a feature reconstruction objective it simplifies the training process while maintaining robustness in handling long-range temporal dependencies. This design allows for more accurate segmentation and object discovery across diverse video domains.



 \begin{figure*}[t!]
    \centerline{
    \includegraphics[width=0.8\linewidth]{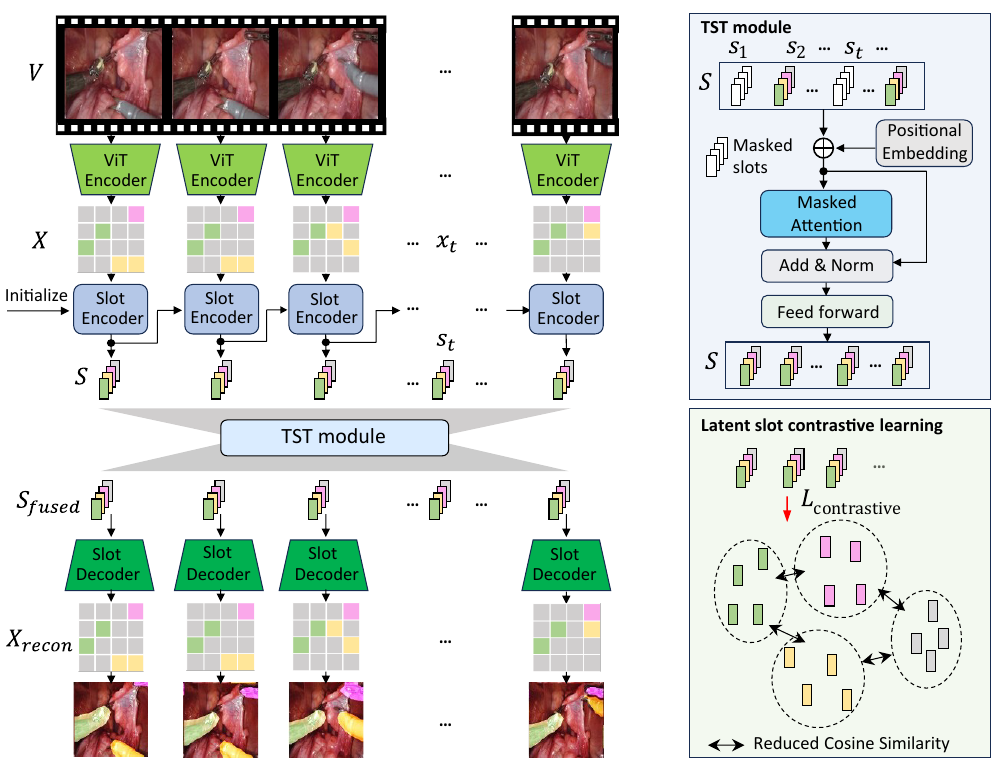}}
    \caption{Overview of our object-centric representation learning framework. Video sequences are encoded into features, processed with a recurrent slot attention mechanism, and refined using a Temporal Slot Transformer (TST). The final slot representations are decoded to reconstruct the input features, with training optimized to minimize reconstruction loss and slot contrastive loss.} 
    \label{fig_method}
\end{figure*}
\section{Methods} 
\label{sec:methods}

%
%

\subsection{Overview}

We base our object-centric representation learning framework (Figure \ref{fig_method}) on slot attention with RNN \citep{kipf2021conditional, singh2022simple,elsayed2022savi++}. A video sequence  $V \in {\mathbb{R}}^{W\times H \times C \times T}$ = $\{I_1,...,I_t,...,I_T\}$, where $I_t \in {\mathbb{R}}^{W\times H\times C}$ is an image at time step $t$ of size $W\times H$ with $C$ channels, and $T$ is a fixed length of the video clip that our model operates on, is encoded using a self-supervised feature extraction model to obtain $X\in {\mathbb{R}}^{N \times D_{feature}\times T}$. Following a recurrent iterative attention approach we obtain the slots representation  $S_{initial}\in {\mathbb{R}}^{K\times d_{slot} \times T}$ = $\{s_1,...,s_t,...,s_T\}$, where $s_t \in {\mathbb{R}}^{K\times d_{slot}}$ are the latent space slots that embed objectness information of image $I_t$. A \gls{tst} module feeds $S_{initial}$ to a masked transformer encoder which allows elements  $s_t$ to attend to each other and aggregate the long-distance information to form $S_{final}$. An image decoder then recurrently maps each element $s_t$ of $S_{final}$ to the video encoding space  $X_{recon} \in {\mathbb{R}}^{N\times D_{feature}\times T}$. The main training objective is to minimize the distance between the reconstructed feature and the original feature representation $X$.

\subsection{Temporal recurrence of object-centric slot encoder}
\label{subsec:methods_encoding_scheme}
 
We encode patches of size $P\times P$ from individual input frames using the \gls{vit} encoder,  producing a stack of patch feature embeddings $X\in {\mathbb{R}}^{N \times D_{feature} \times T}$, where $N=w\times h$ and $(h, w) =(H/P, W/P)$.

Features $x\in {\mathbb{R}}^{N \times D_{feature}}$ representing one individual image are grouped by the slot attention module $f_{\mathrm{SA}}$~\citep{locatello2020object} into $K$ spatial groupings through iterative competitive attention and encoded into $K$ slots $s\in {\mathbb{R}}^{K \times d_{slot}}$. Let $W_k, W_v$ denote the key and value transformation matrices acting on $x$, and $W_q$ the query transformation matrix acting on $s$.  The iterative slot update function is based on dot attention:
\begin{equation} 
s^{\,i+1} := f_{\mathrm{SA}}(x, s^{\,i}) = \hat{A} v 
\end{equation} 

\begin{equation}
{\hat{A}_{mn} := \frac{A_{mn}}{\sum_{l=1}^N A_{ml}}}
\end{equation} 

\begin{equation}
A := \text{softmax} \left( \frac{q k^T}{\sqrt{d}} \right) \in \mathbb{R}^{K \times N}
\end{equation} 
 
where $q = W_q s^{\,i} \in \mathbb{R}^{K \times d_{slot}}$ is the query vector, $k = W_k x \in \mathbb{R}^{N \times D_{feature}}$ is the key vector, $v = W_v x \in \mathbb{R}^{N \times D_{feature}}$ is the value vector, and $A \in \mathbb{R}^{K \times N}$ is the attention matrix. {$\hat{A}\in\mathbb{R}^{K\times N}$ is the normalized attention used for aggregation,, $m \in \{1,\dots,K\}$ is the slots index, $n \in \{1,\dots,N\}$ indexes the input features, and $l \in \{1,\dots,N\}$ is the summation index over features.} 


Slot attention can be seen as a version of self-attention \citep{vaswani2017attention} constrained by an object-centric inductive bias that is much more computationally efficient as the attention query is much smaller ( $K \ll N$, e.g. $N=784$ for patches, $K=5 \sim 20$ for objects in an image). 


The assignment of slots across video frames is initialized by a RNN-like computation where slots $s_t^{\,i=0}$ =$\hat{s}_{t-1}$ for $t$-th frame are initialized by the previous frame's final slot estimation. First frame's slots $s_0^{\,i=0}$ are initialized randomly by sampling from a standard Gaussian. In this way, the permutation of groupings of objects across slots is likely to remain the same at neighboring frames. However, this assignment could not guarantee temporal consistency for longer video sequences. We thus introduce the \gls{tst} module to enhance long-range bidirectional reasoning along the temporal dimension.

\subsection{ Temporal slot transformer}

\label{subsec:methods_temporal_slot_transformer}

The \gls{tst} module facilitates interactions across time steps and enables robust temporal consistency by leveraging positional embeddings and masked training. It processes a sequence of slots $S \in \mathbb{R}^{K\times d_{slot} \times T}$, where $T$ is the number of frames, $K$ is the number of slots, and $d_{slot}$ is the slot dimension. The module comprises the following components:

\paragraph{Temporal Positional Embeddings}  A learnable positional embedding $P_{temporal} \in \mathbb{R}^{  1 \times d_{slot} \times T }$ is added to the slots to encode temporal information:
   \begin{equation}
   S_{pos} = S + P_{temporal}.
   \end{equation}

\paragraph{Masked Transformer Encoder} To promote robust learning, and allow \gls{tst} to have bidirectional reasoning ability (i.e., can predict previous and next slots) we employ random masking during training. For a given masking ratio $\gamma$, a subset of frames is masked by setting their corresponding slot values to zero:
   \begin{equation}
   S_{masked} = S_{pos} \odot M_{slot}, \quad M_{slot} \in \{0, 1\}^{1 \times 1 \times T},
   \end{equation}
   where $M_{slot}$ is the binary mask, and $\odot$ represents element-wise multiplication. The masked slots are then processed through a Multi-head Transformer encoder:
   \begin{equation}
   S_{fused} = \text{TransformerEncoder}(S_{masked},M_{slot}),
   \end{equation}
 Our masked transformer implementation follows previous work in language modeling \citep{kenton2019bert}. {The key to \textit{bi-directionality} is that the self-attention mechanism is non-causal, in contrast to RNN-like models. Self-attention computes dependencies across all time steps simultaneously, allowing each slot to attend to both past and future frames. During training, the model is tasked with reconstructing the original $S_{pos}$ from $S_{masked}$, which encourages learning temporally coherent representations that exploit information from the entire sequence in both directions. This design explicitly enables reasoning in both temporal directions.} 
 
 We use 3 layers of transformer encoder, each with multi-head self-attention ($n_{heads}$=8), feed-forward layers with hidden dimension 4$ \times d_{slot}$ and position embedding. The transformer computes attention across all input slots and masked slots $M_{slot}$, thus modeling dependencies and interactions between unmasked slots effectively. It results in assignments of temporally fused slots $S_{fused} \in \mathbb{R}^{ K \times d_{slot}\times T}$.
By leveraging temporal positional embeddings and masking, the temporal slot transformer achieves robust temporal alignment and enhances long-range temporal reasoning. In addition, it can also serve as a future slot prediction module.
 
\subsection{Slot decoder}
%
As our training objective is feature reconstruction of individual frames, we apply the decoder to all slots in $S$ at image level to obtain the reconstructed video features $X_{recon}$ = $\{x_1,x_2,...,x_T\}$.
We experimented with two types of slot decoders.

\paragraph{MLP broadcast decoder} A simple MLP broadcast decoder \citep{watters2019spatial} has been used in VAE models and previous slot attention research \citep{seitzer2022bridging}. Each slot among the $K$ slots is broadcast to match the number of spatial patches resulting in $N$ tokens for each slot. A learnable positional encoding is added to each token. These tokens are then processed independently using a shared MLP to output reconstructed features $\hat{x}_k$ and associated alpha masks $\alpha_k$ indicating the slot's attentive region.  The final reconstruction $x \in \mathbb{R}^{N \times D_{\text{feature}}}$ is obtained by a weighted sum:
\begin{equation}
x = \sum_{k=1}^K \hat{x}_k \odot m_k\ , \quad m_k = \text{softmax}_k(\alpha_k)
\end{equation} 
where $\odot$ denotes element-wise multiplication. The advantage of this simple design is its efficiency: as the MLP is shared across slots and positions, $m_k$ produced by the decoding is directly deployed as object segmentation masks. 

\paragraph{SlotMixer decoder} The recently introduced SlotMixer
decoder \citep{sajjadi2022object} for 3D object-centric learning has a constant overhead in the number of slots, as it only decodes once per output, requiring less computation. It employs an attention-based approach and operates in three key steps: slots allocation, mixing, and rendering \citep{sajjadi2022object}. The \textit{allocation} step takes as input the slots $s_t \in \mathbb{R}^{K \times d_{slots}}$ and outputs an embedding vector $f \in \mathbb{R}^{N \times d_{slots}}$ using a cross-attention transformer. 
The \textit{mixing} step is similar to a single-head attention step, using the embedding $f$ as queries and the slots $s_t$ as keys, to form an attention map $A_{mix}  \in \mathbb{R}^{N\times K}$, and the slot mix $m$ is obtained as
\begin{equation}
m = s_t A_{mix} \ , \quad m \in \mathbb{R}^{N \times d_{slot}}\enspace .
\end{equation} 
Finally the \textit{rendering} step uses a MLP with position embedding shared with the allocation step to decode $m$ into $x \in\mathbb{R}^{N \times D_{feature}}$, avoiding the need for a broadcast operation.
The learned attention matrix  $A_{mix}$ is applied as $K$ channel segmentation masks with resolution $N=w \times h$. 

Our experimental results demonstrate that, whether using an MLP broadcast decoder or a Slot-Mixer decoder, our method improves upon the vanilla RNN-based baseline and outperforms other state-of-the-art object-centric methods on the surgical video dataset.
%
\subsection{Slot Contrastive Learning}
\label{method_decoder}
To encourage diversity among slots and reduce redundancy, {one possible approach is to adopt clustering-based metrics such as Normalized Mutual Information for designing the loss function. However,} inspired by the SimCLR framework~\citep{chen2020simple}, we adopt a contrastive learning loss based on cosine similarity {, which is more computationally efficient, directly optimizable during training, and guarantees differentiability}. Unlike SimCLR, which computes similarity for positive pairs, our loss enhances dissimilarity between negative pairs~\citep{radford2021learning}. {Unlike conventional methods that use contrastive loss to learn a single global image representation, we apply it to encourage dissimilarity among slot representations that decompose the image and represent objects across video frames.} Let \(u \in \mathbb{R}^{1 \times d_{\text{slot}}}\) represent the one of $K$ slot vectors in $s_t$ for a given frame. For each slot vector $u_i$, we compute its cosine similarity with all other slots {$u_j$} within the same frame:
\begin{equation}
\text{sim}(u_i, u_j) = \frac{u_i \cdot u_j}{\|\mathbf{s}_i\| \|\mathbf{s}_j\|}.
\end{equation} 
To exclude self-similarity, the cosine similarity matrix \(C \in \mathbb{R}^{K \times K}\) is adjusted by subtracting the identity matrix:
\begin{equation}
C_{ij} = \text{sim}(u_i, u_j) - \delta_{ij},
\end{equation}
where \(\delta_{ij}\) is the Kronecker delta (i.e., \(\delta_{ij} = 1\) if \(i = j\), and \(0\) otherwise).

The total contrastive loss for each pair of slot vectors across \(T\) frames is then computed as:
\begin{equation}
\mathcal{L}_{\text{contrast}} = \frac{1}{T \cdot K^2} \sum_{t=1}^{T} \sum_{i=1}^{K} \sum_{j=1}^{K} \left[ - \log \frac{\exp(-C_{ij} / \tau)}{\sum_{k=1}^{K} \exp(-C_{ik} / \tau)} \right],
\end{equation}
where \(\tau > 0\) is a temperature parameter.  {By minimizing} {t}his loss {the cosine similarity between slots is reduced, which will push the slot vectors to different directions in the representation hyper-space, thus increasing the distance between different slot vectors that now lie on a sphere and point in quasi-orthogonal directions.}

\subsection{Training loss}
The main learning objective is to decode final slots to feature space $X_{recon}$. Reconstruction loss 
\begin{equation}
\mathcal{L}_{\text{recon}} = \|X_{\text{recon}} - X\|_2^2
\end{equation} 

guides the alignment between the original and reconstructed features.

The final loss function combines the reconstruction loss and the contrastive loss:
\begin{equation} 
\mathcal{L}_{\text{final}} = \mathcal{L}_{\text{recon}} + \alpha \mathcal{L}_{\text{contrast}},
\end{equation} 
where \(\alpha\) is a scaling factor to balance the two terms.

The inclusion of the contrastive loss ensures that the slot representations remain diverse, while the reconstruction loss helps align the slots with the input features, guiding effective representation learning.

\section{Experiments and Results}
\label{sec:experiments_results}
\subsection{Dataset}
\label{sec_dataset}
We evaluate the performance of \gls{sbert} on 4 surgical video dataset from 3 types of surgery (Table \ref{tab_data}):

\textit{MICCAI 2022 SurgToolLoc Challenge Data}: This dataset consists of 24,642 30 second video clips at 60 FPS collected from animal, phantom, and simulator-based surgeries \citep{zia2023surgical}. It includes a total of 13 instrument types, with up to four instruments in a clip. For simplicity, we refer to this dataset as MICCAI. For our use we downsample clips to 1 FPS. 

\textit{Cholec80}: The Cholec80 dataset \citep{twinanda2016endonet} contains 80 cholecystectomy surgery videos downsampled to 1 FPS with presence labels for seven surgical tools in each frame. Following the preprocessing steps of a weakly supervised learning study \citep{liao2024disentangling}, we derived a \gls{top} video subset from Cholec80 resulting in 5,296 clips with 30 frames each{, in which objects are not necessarily present in every frame, but can rather transition in or out of the field of view}. The test set is based on 100 clips from the CholecSeg8K dataset \citep{hong2020cholecseg8k}, which contains 8,000 frames with instrument and anatomy segmentation from 17 Cholec80 videos. \DIFadd{To evaluate long-range object-centric tracking under challenging
multi-instrument interaction conditions, we curate a separate test set of
20 unseen cholecystectomy videos, each comprising 30-frame (30\,s) sequences.
These videos feature frequent instrument interactions, occlusions, and
repeated instrument entry and exit events, posing a substantial challenge
for maintaining temporal identity consistency.}

\textit{EndoVis 2017 Robotic Instrument Segmentation Challenge Data}: This dataset contains 2,400 annotated frames from videos of abdominal surgery sampled at 1 FPS. We generated 480 video clips by grouping five consecutive frames per clip. The data includes seven instrument types with a maximum of four instruments present in a single clip.

\textit{Thoracic Robotic Surgery}: This dataset includes data from 40 robot-assisted right upper lobectomies (RULs) for lung malignancy performed at the Toronto General Hospital  between 2014 and 2023. A total of 264 annotated clips were selected for this study.

{Among these four datasets, only the derivative of \textit{Cholec80}, CholecSeg8k provides both instrument and tissue ground truth masks for evaluation, while other datasets provide only instrument masks.}

{To demonstrate the benefit of applying our method to broader domains, We also evaluate Slot-BERT on non-surgical datasets, including a natural real-world video dataset YT-VIS \mbox{\citep{xu2019youtubevis}} and a synthetic multi-object tracking data MOVi-E \mbox{\citep{greff2022kubric}}. Results on these two datasets are concluded in Supplementary material \ref{supple_natural}.}
  
 \begin{table}[t!]
\caption{Data summary. Testing sets are based on frames in annotated clips and the rest of the data is used for training. Except for the Thoracic dataset, all frames in annotated clips have segmentation masks.}
\label{tab_data}
\centering
\setlength\arrayrulewidth{0.9pt}
\setlength\doublerulesep{0.9pt} 
\resizebox{1.0\linewidth}{!}{%
\begin{tabular}{lcccc}
\hline
  & MICCAI & Cholec &  {Endovis} &  Thoracic \\ \hline
  Procedure                   &   abdominal          & cholecystectomy            &   abdominal           &  thoracic  \\          
Number of clips                       & 24642                      & 6300                       & 480                         & 264                          \\
Frames per clip                         & 30                         & 30                         & 5                           & 30                           \\
Annotated clips                        & 100                        & 100                        & 480                         & 264                          \\

Annotated frames                        & 3000                       & 3000                       & 2400                        & 550                          \\
Clips in training set               & 24542                      & 5296                       & ---                          & ---                          \\  \hline 

\end{tabular}
}
\end{table}

\subsection{Metrics  }
  We evaluate our approach based on the quality of the slot masks produced by the decoder using four primary metrics: video foreground ARI (FG-ARI) \citep{greff2019iodine}, video mean best overlap (mBO) \citep{pont2016multiscale}, mean best Hausdorff Distance (mBHD) and CorLoc\citep{bilen2016weakly} for evaluating localization accuracy. FG-ARI, adapted from a widely-used metric in object-centric research, measures the similarity between predicted object masks and ground truth masks, focusing on how effectively objects are segmented. mBO, on the other hand, evaluates the alignment between predicted and ground truth masks using the intersection-over-union (IoU) metric. For mBO, each ground truth mask is matched to the predicted mask with the highest IoU via Hungarian matching, and the average IoU is computed across all matched pairs.
While FG-ARI primarily measures segmentation quality, mBO provides a more comprehensive assessment by including background pixels. Furthermore, the video version of mBO (mBO-V) also accounts for the temporal consistency of masks throughout the video. \DIFadd{To explicitly evaluate long-range temporal consistency, we additionally report identity-aware tracking metrics, including the slot-level IDF1 score~\mbox{\citep{ristani2016performance}} and {Temporal Identity Persistence} (T-IDP)~\mbox{\citep{milan2016mot16}}. IDF1 measures identity precision and recall across frames, while T-IDP quantifies the fraction of frames in which each ground-truth instance is consistently associated with the same predicted slot.}
To facilitate comparisons at the frame level, we include the image-based version of mBO (mBO-F), which is computed on individual frames. This metric evaluates segmentation quality at the image level without considering temporal consistency. Additionally, to assess the boundary accuracy of the predicted masks, we calculate the mean best Hausdorff Distance (mBHD) for the best-overlapping masks produced by the object-centric learning model. CorLoc is also calculated based on overlapped object instance bounding box, given a threshould IoU$>$0.5.

\subsection{Baseline models}
We compare our method with state-of-the-art object-centric algorithms, including SAVi \citep{kipf2021conditional}, STEVE \citep{singh2022simple}, DINOSaur \citep{seitzer2022bridging}, Video-Saur \citep{zadaianchuk2024object}, and Slot-Diffusion \citep{wu2023slotdiffusion}.

SAVi \citep{kipf2021conditional} is a weakly supervised, recurrent video object-centric method that leverages the temporal dynamics of video data through optical flow. It conditions the initial slots on the central coordinates of objects. Since our training data lacks any form of supervision, we adopt a fully unsupervised version of SAVi that excludes initial state conditioning.\footnote{\url{https://github.com/google-research/slot-attention-video/}}

STEVE \citep{singh2022simple} is an unsupervised model designed for object-centric learning in videos. Like SAVi, it employs an RNN-like slot initialization mechanism. However, STEVE replaces the broadcast decoder used in video slot attention with a specialized transformer-based slot decoder conditioned on slots. Its learning objective focuses on reconstructing individual video frames.\footnote{\url{https://github.com/singhgautam/steve}}

DINOSaur \citep{seitzer2022bridging} bridges the slot attention algorithm to real-world images by replacing the traditional image reconstruction objective with feature reconstruction. It can be implemented with feature encoders such as DINO, MAE, MoCo-v3 \citep{chen2021empirical}, or MSN \citep{assran2022masked}, and the encoder can be either Convolutional Neural Networks (CNNs) or Vision Transformers (ViTs). For our comparisons, we use the version based on the DINO ViT feature extractor.\footnote{\url{https://github.com/amazon-science/object-centric-learning-framework}}

Video-Saur \citep{zadaianchuk2024object}, is a recurrent video object-centric method that replaces feature map reconstruction with next-frame feature cosine similarity as its reconstruction objective. It also employs a video-specific slot-mixer decoder for object-centric learning. Following the official implementation of Video-Saur, we report results using a combination of feature reconstruction and similarity reconstruction, with the DINO ViT as the feature extractor.\footnote{\url{https://github.com/martius-lab/videosaur}}

Slot-Diffusion \citep{wu2023slotdiffusion} is an object-centric Latent Diffusion Model (LDM) designed for both
image and video data. It introduces the replace of the decoding module for object-centric learning with slot-conditioned diffusion generation models. Based on the video version of Slot-Diffusion, similar to the learning objective of other diffusion models \citep{rombach2022high}, we trained it to reconstruct images of surgical videos following its encoding and decoding pipeline\footnote{\url{https://github.com/Wuziyi616/SlotDiffusion}}.

\subsection{Experiment setup}
\label{sec:experiment_setup}
 All raw images of video from different datasets were first cropped to remove zero pixels and then resized to 224$\times$224. For evaluating segmentation masks, they are upscaled to the same size for computing evaluation metrics. Our training and inference are implemented on a workstation with 503 GB ram based NVIDIA RTX 6000 Ada Generation GPUs. Adam optimizer with a learning rate of $1\times10^{-4}$ and a weight decay of $1\times10^{-5}$ was used for training.   All models are trained under batch size of 4.
 
We first trained the models on the MICCAI and Cholec datasets. Training duration was 80 epochs for MICCAI and 100 epochs for Cholec. Performance of these models was then tested on all annotated frames from the MICCAI and Cholec testing sets. Next, we tested performance in an unsupervised transfer learning scenario. We used the MICCAI pre-trained weights as the original model, fine-tuned them on the Cholec training data for 10 epochs and measured performance on the Cholec testing set. To measure how well the learned object-centric representations generalize to novel, unseen databases without additional training we measured zero-shot segmentation performance by using the MICCAI trained models to segment testing sets in the EndoVis, Cholec, and Thoracic datasets. 

\DIFdel{All models were trained with the first 5 frames of each clip, even when the clips had longer sequences.} For testing, \DIFdel{each} \DIFdel{Most} videos were  processed by the trained model three times to calculate an average score and standard deviation, ensuring that any potential instabilities in the model's outputs were accounted for~\citep{wu2023slotdiffusion}. \DIFadd{Five processing repetitions were used in ablation studies.}

\DIFdel{Finally, we tested performance on longer frame sequences of 7, 11, and 30 frames and evaluated a variation of our model on longer sequences by duplicating the \gls{tst} module as a \textit{next-slot initializer}.} \DIFadd{For testing sequences longer than the training context window, we adopt an online recurrent inference strategy with a sliding window. In addition, we evaluate a variant of our model that extends temporal reasoning to longer sequences by duplicating the \gls{tst} module and using it as a \textit{next-slot initializer}.}
 
\subsection{Unsupervised segmentation performance} 

\begin{table*}[t!]
\caption{Unsupervised training from scratch. Bold values indicate the best performance for each dataset.}
\label{tab_scratch}
\centering
\setlength\arrayrulewidth{0.9pt}
\setlength\doublerulesep{0.9pt} 
\resizebox{1.0\linewidth}{!}{%
\begin{tabular}{clllllll}
\hline
Datasets                  &  & Method         & mBO-V (\%)            & mBO-F (\%)            & mBHD (↓)              & FG-ARI (\%)    & CorLoc  (\%)     \\ \hline
\multirow{6}{*}{MICCAI}   &  & DINO-Saur\citep{seitzer2022bridging} & 38.2 ± 0.9          & 42.9 ± 0.5          & 62.5 ± 1.3            & 48.4 ± 0.7  & 45.8 ± 2.2           \\ 
                          &  & SAVi\citep{kipf2021conditional}     & 29.4 ± 0.2          & 33.2 ± 0.1          & 81.7 ± 1.0            & 36.6 ± 0.2  & 40.0 ± 0.5           \\ 
                          &  & STEVE\citep{singh2022simple}        & 27.9 ± 0.2          & 31.5 ± 0.1          & 139.9 ± 0.6           & 34.3 ± 0.1  & 17.0 ± 0.4           \\ 
                          &  & Slot-Diffusion\citep{wu2023slotdiffusion} & 37.5 ± 0.1          & 42.2 ± 0.1          & 70.5 ± 0.2            & 46.3 ± 0.0  & 42.0 ± 0.2           \\ 
                          &  & Video-Saur\citep{zadaianchuk2024object}   & 46.3 ± 0.4          & 50.1 ± 0.4          & 53.9 ± 1.3            & 55.1 ± 0.5  & 60.1 ± 1.9           \\ 
                          &  & Ours           & \textbf{48.9 ± 0.2} & \textbf{52.8 ± 0.3} & \textbf{44.2 ± 0.6}   & \textbf{58.2 ± 0.2} & \textbf{70.7 ± 0.8} \\ \hline
 
\multirow{6}{*}{Cholec}   &  & DINO-Saur\citep{seitzer2022bridging} & 25.7 ± 1.0          & 25.5 ± 0.5          & 75.9 ± 0.6            & 33.9 ± 0.7  & 29.1 ± 0.5           \\ 
                          &  & SAVi\citep{kipf2021conditional}     & 18.9 ± 0.2          & 18.1 ± 0.1          & 107.0 ± 0.2           & 23.7 ± 0.1  & 15.3 ± 0.1           \\ 
                          &  & STEVE\citep{singh2022simple}        & 19.5 ± 0.1          & 18.7 ± 0.0          & 108.9 ± 0.2           & 24.2 ± 0.0  & 19.8 ± 0.7           \\ 
                          &  & Slot-Diffusion\citep{wu2023slotdiffusion} & 12.8 ± 0.0          & 15.7 ± 0.0          & 91.4 ± 0.3            & 19.4 ± 0.0  & 18.9 ± 0.2           \\ 
                          &  & Video-Saur\citep{zadaianchuk2024object}   & 26.1 ± 0.3          & 25.8 ± 0.2          & 74.9 ± 0.3            & 34.2 ± 0.2  & 34.6 ± 1.9           \\ 
                          &  & Ours           & \textbf{28.8 ± 0.3} & \textbf{27.8 ± 0.4} & \textbf{64.8 ± 0.9}   & \textbf{37.0 ± 0.6} & \textbf{35.3 ± 2.0} \\ \hline

\multirow{6}{*}{{Cholec + Tissue Mask}} 
                          &  & DINO-Saur\citep{seitzer2022bridging} & 30.2 ± 0.1 & 36.7 ± 0.5 & 90.9 ± 1.0  & 41.4 ± 0.6 & 22.5 ± 0.2 \\
                          &  & SAVi\citep{kipf2021conditional}     & 27.8 ± 0.0 & 29.3 ± 0.0 & 117.2 ± 0.1 & 35.4 ± 0.1 & 16.4 ± 0.0 \\
                          &  & STEVE\citep{singh2022simple}        & 28.5 ± 0.0 & 33.6 ± 0.1 & 127.7 ± 0.0 & 33.0 ± 0.0 & 16.5 ± 0.1 \\
                          &  & Slot-Diffusion\citep{wu2023slotdiffusion} & 29.3 ± 0.2 & 34.0 ± 0.2 & 99.1 ± 0.2  & 35.5 ± 0.2 & 23.3 ± 0.2 \\
                          &  & Video-Saur\citep{zadaianchuk2024object}   & 30.9 ± 0.1 & 33.7 ± 0.1 & 90.3 ± 0.1  & 43.1 ± 0.3 & 24.8 ± 0.3 \\
                          &  & Slot-BERT (Ours)                     & \textbf{33.4 ± 0.5} & \textbf{40.2 ± 0.3} & \textbf{85.4 ± 0.8} & \textbf{44.1 ± 0.5} & \textbf{25.9 ± 0.3} \\ \hline

\end{tabular}
}

\begin{tablenotes}
\footnotesize
\item { $^\dagger$ We report Cholec results on instruments only (Cholec) for easier comparison to MICCAI, and on both instruments and tissue (Cholec + Tissue Masks).}
 
\end{tablenotes}
\end{table*}

This section presents the performance of models trained from scratch on MICCAI and Cholec data. Table~\ref{tab_scratch} highlights the effectiveness of our approach across multiple evaluation metrics, demonstrating its robustness to various dataset characteristics.

All methods achieve their highest performance on the MICCAI dataset, reflecting the advantages of having a larger collection of data. Our method surpasses all competing methods in every metric:~mBO-V (48.90\%; +2.60\% increase), mBO-F (52.80\%; +2.70\% increase), mBHD (44.2; -9.7 decrease), FG-ARI (58.20\%; 3.10\% increase) and CorLoc (70.7\%; 10.6\% increase). These improvements suggest that the model benefits from the dataset’s scale and diversity. Compared to the closest competitor, Video-Saur, our method  significanly improves mBO-V, mBHD and CorLoc metrics, indicating better segment boundary and instance accuracy.

\begin{figure}[t!]
    \centerline{
    \includegraphics[width=1.0\linewidth]{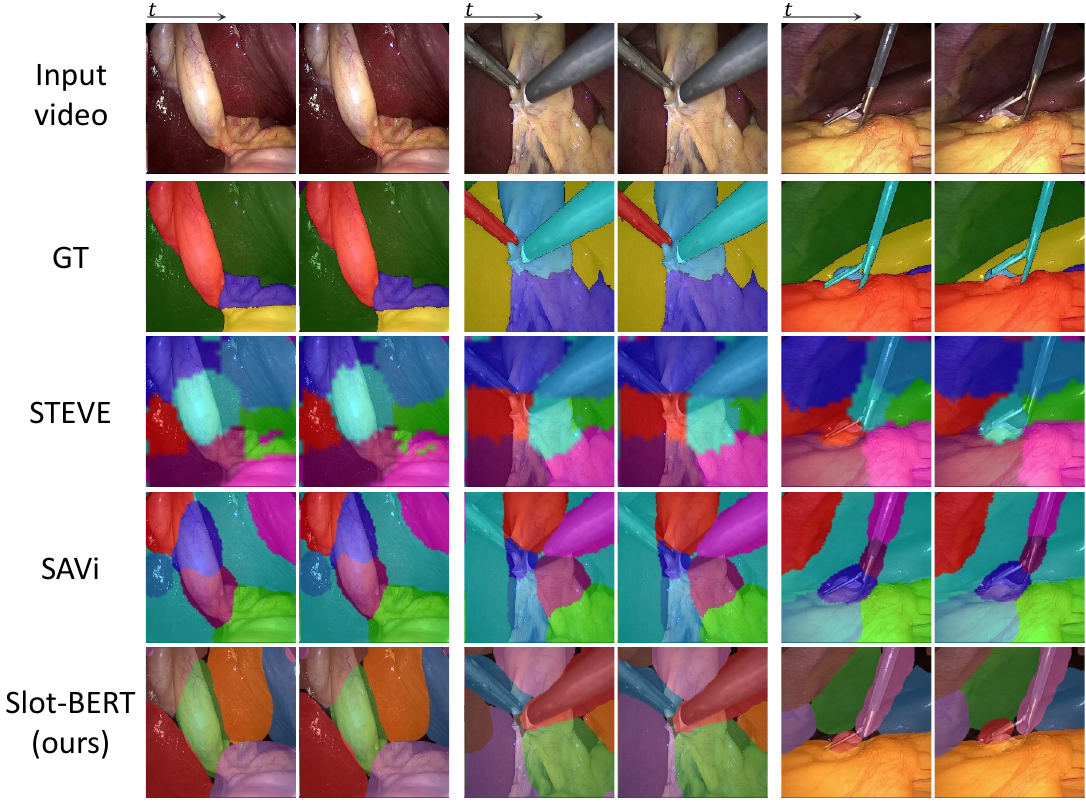}}
    \caption{{Qualitative results of unsupervised segmentation using STEVE, SAVi and our method on the Cholec dataset, which can be evaluated both tissue and instrument ground truth (GT). Slot-BERT demonstrates a stronger ability to delineate instruments from tissue and to separate tissues with different textures.  }} 
    \label{fig_tissue}
\end{figure}

The performance on the Cholec dataset is comparatively lower for all models. Nevertheless, our method outperforms all baselines, with notable gains over Video-Saur  (e.g., +2.7\% in MBO-V, +2.8\% in FG-ARI and -10.1 in mBHD {on instrument segmentation; +2.7\% in MBO-V when evaluated with tissue masks }). It is evident that the smaller size of the dataset and less diversity in the data impact performance compared to MICCAI, the latter being almost five times larger than Cholec. An additional challenge in Cholec dataset is the frequent instrument disappearance/reappearance~\citep{liao2024disentangling} and the small size of some objects which pose a challenge because the slot number is fixed. These data characteristics also result in the phenomenon where mBO-F is smaller than mBO-V, unlike in MICCAI data, where most methods achieve a higher mBO-F instead. {Figure \ref{fig_tissue} presents qualitative results of unsupervised segmentation, illustrating our method’s advantage in delineating instruments from tissue and discriminating tissue with different textures compared to STEVE and SAVi.}

\subsection{Transfer learning performance}

  \begin{table}[t!]
\caption{Transfer learning performance of Slot-BERT compared to object-centric SOTA methods. Fine-tuning of the model trained on the larger MICCAI dataset leads to better performance in comparison to models trained from scratch on Cholec.}
\label{tab_transfer}
\centering
\setlength\arrayrulewidth{0.9pt}
\setlength\doublerulesep{0.9pt} 
\resizebox{1.0\linewidth}{!}{%
\begin{tabular}{llll}
\hline
     Methods                                                      & \multicolumn{1}{c}{mBO-V (\%)} & \multicolumn{1}{c}{mBO-F (\%)} & \multicolumn{1}{c}{FG-ARI (\%)} \\ \hline
DINO-Saur\citep{seitzer2022bridging}      & 26.3 ± 1.0              & 26.0 ± 0.9              & 34.7 ± 1.1              \\ 
SAVi\citep{kipf2021conditional}           & 18.2 ± 0.1              & 17.6 ± 0.0              & 23.0 ± 0.1              \\ 
STEVE\citep{singh2022simple}              & 17.4 ± 0.0              & 16.7 ± 0.0              & 22.1 ± 0.1              \\ 
Slot-Diffusion\citep{wu2023slotdiffusion} & 23.4 ± 0.1              & 22.9 ± 0.1              & 29.8 ± 0.1              \\ 
Video-Saur\citep{zadaianchuk2024object}   & 28.7 ± 0.5              & 28.3 ± 0.5              & 37.3 ± 0.6              \\ 
\multicolumn{1}{l}{Ours}                                   & \textbf{31.4 ± 0.6}    & \textbf{30.1 ± 0.3}    & \textbf{40.0 ± 0.7}       \\ \hline
\end{tabular}
}
\end{table}

We evaluate the transfer learning performance of Slot-BERT by taking the model trained on the MICCAI dataset and fine-tuning it on the Cholec dataset. We fine-tune the model for 10 epochs. Our experiments show that fine-tuning provides significant performance gains over training the model from scratch on the Cholec dataset indicating that object specific representations benefit from training on larger datasets and can be easily reused on smaller datasets.

A comparison of transfer learning performance with SOTA is summarized in Table \ref{tab_transfer}. We observe that Slot-BERT outperforms all other methods in all three metrics, outperforming Video-Saur by 2.7\% in mBO-V and FG-ARI. Fine-tuning results in an increase of 2.6\% in \textit{mBO-V}, 2.2\% in \textit{mBO-F}, and 3.0\% in \textit{FG-ARI} compared to training on Cholec data, confirming the plausibility of transfer learning to leverage large databases and enhance the performance of object-centric models on new datasets.

\subsection{Zero-shot  performance }
\label{result_0_transfer}

Table \ref{tab_zeroshot} illustrates the performance of models trained on MICCAI and tested directly on unseen datasets without additional supervision. When comparing the zero-shot results to models trained fully from scratch on the respective testing domains (refer to Table \ref{tab_scratch}), our method demonstrates performance that is highly competitive with domain-specific training, achieving almost identical results on the Cholec database and commendable results on the new EndoVis and Thoracic datasets. 

\begin{figure*}[t!]
    \centerline{
    \includegraphics[width=1.0\linewidth]{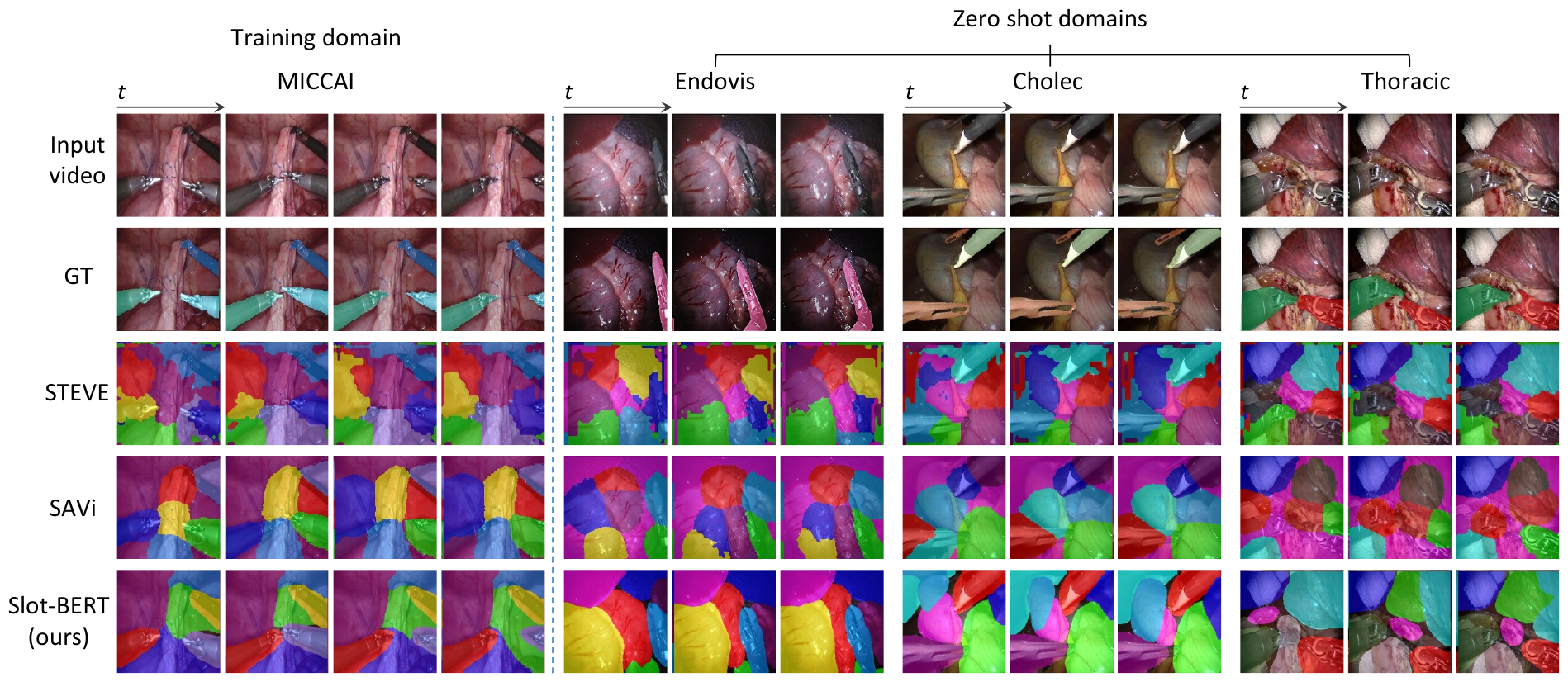}}
    \caption{Qualitative results of zero-shot experiments on unseen datasets using our method compared to STEVE and SAVi. Slot-BERT demonstrates superior adaptability, successfully segmenting unseen surgical instruments, while alternative methods exhibit degraded performance. Despite limitations in precise boundary detection due to patch-based processing, our method achieves high object localization coverage.} 
    \label{fig_zeroshot}
\end{figure*}

 Qualitative results of the zero-shot performance, illustrated in Figure \ref{fig_zeroshot}, show that our method successfully segments unseen surgical instruments in new surgical scenes. In contrast, methods like STEVE and SAVi exhibit degraded adaptability, failing to capture objectiveness or accurately locate instruments in unseen videos.

As the Thoracic dataset is sparsely annotated (approximately two frames per clip), the mBO-V is less relevant than mBO-F. Looking at the results for the Thoracic dataset where frame lengths differ significantly compared to Endovis and Cholec, our model still outperforms others, achieving an 2.2\% better mBO-F and 6.4 decrease in mBHD compared to Video-Saur. These results validate the robustness of our model in diverse settings.

Figure \ref{fig_zeroshot} also reveals a drawback of our approach for the accuracy of segmentation masks. As the frames are divided into patches, our method is not accurate at detecting precise object boundaries, in particular for instruments, as the immediate background of instruments is highly correlated across scenes. The method tends to however correctly locate the whole object, leading to a low false negative rate.

\begin{table*}[t!]
\definecolor{customgray}{rgb}{0.5, 0.5, 0.5} 

\caption{Performance comparison in zero-shot segmentation. Zero-shot results demonstrate the transferability of models pre-trained on MICCAI data when applied to unseen domains. Results with the best mean values are highlighted in bold.}
\label{tab_zeroshot}
\centering
\setlength\arrayrulewidth{0.9pt}
\setlength\doublerulesep{0.9pt} 
\begin{tabularx}{0.90\linewidth}{clllll}
\hline
Dataset                   & Method                                                     & \multicolumn{1}{c}{mBO-V (\%)} & \multicolumn{1}{c}{mBO-F (\%)} & \multicolumn{1}{c}{mBHD (↓)} & \multicolumn{1}{c}{FG-ARI (\%)} \\ \hline
\multirow{6}{*}{Endovis}  & DINO-Saur\citep{seitzer2022bridging}      & 34.1 ± 0.3            & 38.7 ± 0.4             & 64.4 ± 1.1               & 45.4 ± 0.5              \\  
                          & SAVi\citep{kipf2021conditional}           & 28.2 ± 0.2            & 32.4 ± 0.1             & 81.6 ± 0.2             & 36.9 ± 0.0              \\  
                          & STEVE\citep{singh2022simple}              & 26.0 ± 0.1            & 29.8 ± 0.1             & 143.7 ± 0.2             & 33.9 ± 0.1              \\  
                          & Slot-Diffusion\citep{wu2023slotdiffusion} & 33.3 ± 0.1            & 35.9 ± 0.1             & 95.6 ± 0.1             & 40.6 ± 0.1              \\  
                          & Video-Saur\citep{zadaianchuk2024object}   & 42.7 ± 0.2            & 46.7 ± 0.2             & 56.2 ± 0.2             & 53.5 ± 0.1              \\  
                          & Ours                                                       & \textbf{43.5 ± 0.3}    & \textbf{47.6 ± 0.3}    & \textbf{50.7 ± 0.4}    & \textbf{54.4 ± 0.3}     \\ \hline
\multirow{6}{*}{Thoracic $^\dagger$} & DINO-Saur\citep{seitzer2022bridging}      & 27.7 ± 0.6            & 34.7 ± 0.5             & 90.7 ± 0.7             & 28.2 ± 0.5              \\  
                          & SAVi\citep{kipf2021conditional}           & 24.9 ± 0.2            & 28.5 ± 0.0             & 105.3 ± 1.0            & 21.7 ± 0.2              \\  
                          & STEVE\citep{singh2022simple}              & 24.0 ± 0.0            & 30.2 ± 0.1             & 127.9 ± 0.6            & 22.6 ± 0.0              \\  
                          & Slot-Diffusion\citep{wu2023slotdiffusion} & 29.3 ± 0.1            & 36.9 ± 0.1             & 103.1 ± 0.2            & 28.1 ± 0.1              \\  
                          & Video-Saur\citep{zadaianchuk2024object}   & 36.7 ± 0.1            & 48.9 ± 0.1             & 72.8 ± 0.4             & 39.2 ± 0.1              \\  
                          & Ours                                                       & \textbf{37.7 ± 0.2}    & \textbf{51.1 ± 0.3}    & \textbf{66.4 ± 0.3}    & \textbf{40.5 ± 0.1}     \\ \hline
\multirow{6}{*}{Cholec}   & DINO-Saur\citep{seitzer2022bridging}      & 21.5 ± 0.8            & 20.7 ± 0.4             & 82.8 ± 1.3             & 28.1 ± 0.9              \\  
                          & SAVi\citep{kipf2021conditional}           & 17.8 ± 0.1            & 17.3 ± 0.1             & 106.1 ± 0.7            & 22.6 ± 0.1              \\  
                          & STEVE\citep{singh2022simple}              & 17.3 ± 0.1            & 16.6 ± 0.1             & 139.2 ± 0.9            & 21.9 ± 0.1              \\  
                          & Slot-Diffusion\citep{wu2023slotdiffusion} & 17.8 ± 0.1            & 17.0 ± 0.0             & 106.1 ± 0.3            & 20.9 ± 0.1              \\  
                          & Video-Saur\citep{zadaianchuk2024object}   & 27.0 ± 0.5            & 26.1 ± 0.5             & 72.5 ± 1.5             & 35.5 ± 0.6              \\  
                          & Ours                                                       & \textbf{29.2 ± 0.1}    & \textbf{27.9 ± 0.1}    & \textbf{64.0 ± 0.6}    & \textbf{37.7 ± 0.0}     \\ \hline
\end{tabularx}

\begin{tablenotes}
\footnotesize
\item $^\dagger$When zero-shot on thoracic the video length is increased to 30 frames in comparison to Endovis and Cholec (5 frames).
 
\end{tablenotes}
\end{table*}

%
%

\subsection{\DIFdel{Adaptation to l} Longer \DIFdel{sequences} \DIFadd{ and more challenging sequences} }
\label{sec_longer}
\begin{figure}[t!]
    \centerline{
    \includegraphics[width=1.0\linewidth]{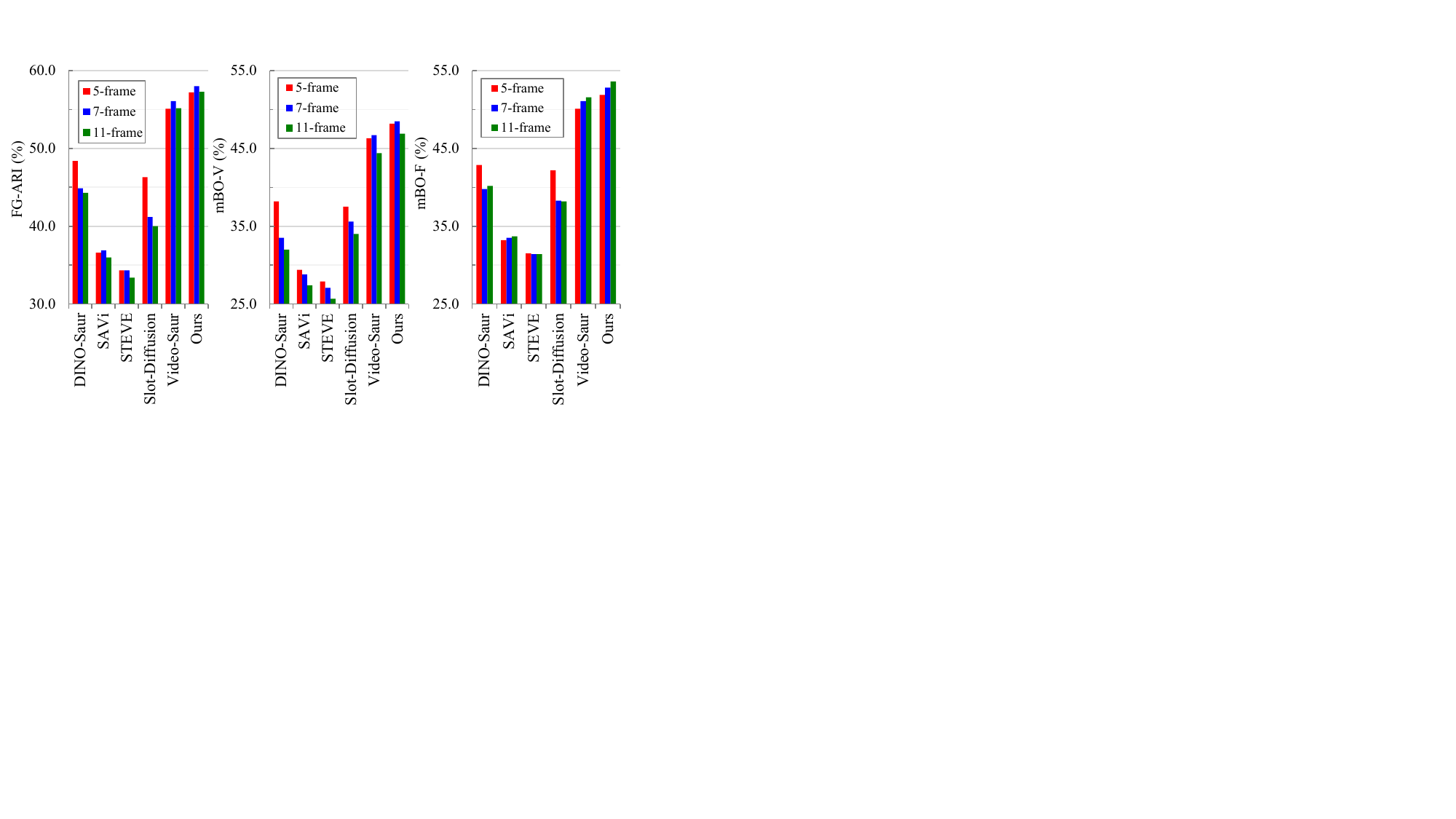}}
    \caption{Impact of sequence length on performance. The plot illustrates mBO-F and mBO-V metrics across different sequence lengths for various methods. Our method demonstrates stronger temporal consistency, with minimal degradation in video-level accuracy (mBO-V) as sequence length increases, outperforming SOTA approaches across all evaluated settings.} 
    \label{fig_longer}
\end{figure}

\begin{figure}[t!]
    \centerline{
    \includegraphics[width=0.95\linewidth]{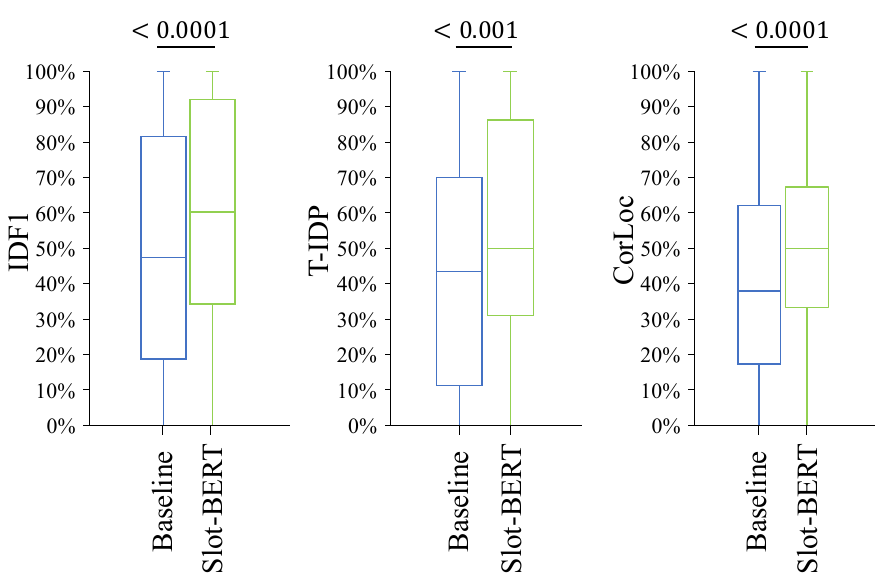}}
    \caption{ \DIFadd{Box plot comparison of Slot-BERT and the baseline in a challenging scenario characterized by frequent object entry and exit. Long-range temporal tracking performance is assessed using IDF1 and Temporal Identity Persistence (T-IDP), together with object localization accuracy (CorLoc). } } 
    \label{fig_tracking}
\end{figure}

\begin{table*}[t!]
\caption{Comparison of performance metrics for sequences of 7 and 11 frames. Metrics include mBO-V, mBO-F, mBHD, and FG-ARI. Our method consistently achieves the highest scores across all metrics, with further improvements observed when incorporating the future slot prediction mechanism, showcasing robustness and adaptability to longer sequences. Results with the best mean values are highlighted in bold.}
\label{tab_longer}
\centering
\setlength\arrayrulewidth{0.9pt}
\setlength\doublerulesep{0.9pt} 
\resizebox{1.0\linewidth}{!}{%
\begin{tabular}{lllllllllll}
\hline
                                                           & \multicolumn{5}{c}{7 Frames}                                                                                         &  & \multicolumn{4}{c}{11 Frames}                                                                      \\ \cline{2-6} \cline{8-11} 
Method                                                     & mBO-V (\%)             & mBO-F (\%)             & mBHD (↓)               & \multicolumn{2}{l}{FG-ARI (\%)}                &  & mBO-V (\%)             & mBO-F (\%)             & mBHD (↓)                & FG-ARI (\%)            \\ \hline
DINO-Saur\citep{seitzer2022bridging}                       & 33.50 ± 0.90           & 39.80 ± 0.20          & 64.0 ± 1.9            & \multicolumn{2}{l}{44.90 ± 0.30}              &  & 32.00 ± 0.30           & 40.20 ± 0.40          & 62.8 ± 1.1            & 44.30 ± 0.40          \\ 
SAVi\citep{kipf2021conditional}                           & 28.80 ± 0.30           & 33.50 ± 0.20          & 80.9 ± 0.9            & \multicolumn{2}{l}{36.90 ± 0.20}              &  & 27.40 ± 0.20          & 33.70 ± 0.30          & 81.7 ± 0.5            & 36.00 ± 0.30          \\ 
STEVE\citep{singh2022simple}                              & 27.10 ± 0.10           & 31.40 ± 0.10          & 141.0 ± 0.7           & \multicolumn{2}{l}{34.30 ± 0.10}              &  & 25.70 ± 0.00           & 31.40 ± 0.00          & 140.3 ± 0.7           & 33.40 ± 0.00          \\ 
Slot-Diffusion\citep{wu2023slotdiffusion}                 & 35.60 ± 0.20           & 38.30 ± 0.10          & 79.6 ± 0.1            & \multicolumn{2}{l}{41.20 ± 0.10}              &  & 34.00 ± 0.10           & 38.20 ± 0.10          & 80.2 ± 0.2            & 40.00 ± 0.10          \\ 
Video-Saur\citep{zadaianchuk2024object}                   & 46.70 ± 0.30           & 51.10 ± 0.20          & 52.5 ± 0.9            & \multicolumn{2}{l}{56.10 ± 0.30}              &  & 44.40 ± 0.30           & 51.60 ± 0.20          & 51.5 ± 0.4            & 55.20 ± 0.20          \\ 
Ours                                                       & {48.00 ± 0.50} & {52.30 ± 0.30} & \textbf{44.1 ± 0.7}   & \multicolumn{2}{l}{{57.50 ± 0.50}}      &  & {46.20 ± 0.50} & {53.10 ± 0.30} & {43.8 ± 0.9}    & {56.80 ± 0.40} \\ 
Ours + Future slot prediction                              & \textbf{48.50 ± 0.80} & \textbf{52.80 ± 0.70} &  {44.14 ± 1.1} & \multicolumn{2}{l}{\textbf{58.00 ± 0.90}}      &  & \textbf{46.90 ± 0.00} & \textbf{53.60 ± 0.10} & \textbf{43.10 ± 0.16}  & \textbf{57.30 ± 0.30} \\ \hline
\end{tabular}
}
\end{table*}

\begin{table}[t!]
\caption{{Comparison of performance metrics (mBO-V, mBO-F, FG-ARI, and Corloc) video for sequences of 30 seconds. Our method achieves the best results across all metrics, demonstrating robustness and adaptability to long sequences. Results with the best mean values are highlighted in bold.}}
\label{tab_30s}
\centering
\setlength\arrayrulewidth{0.9pt}
\setlength\doublerulesep{0.9pt} 
\resizebox{1.0\linewidth}{!}{%
\begin{tabular}{lcccc}
\hline
Method & mBO-V (\%) & mBO-F (\%) & FG-ARI (\%) & Corloc (\%) \\ \hline
DINO-Saur\citep{seitzer2022bridging} & 30.96 ± 0.29 & 42.15 ± 0.09 & 45.11 ± 0.41 & 52.79 ± 0.71 \\ 
SAVi\citep{kipf2021conditional}     & 25.39 ± 0.25 & 35.23 ± 0.02 & 35.98 ± 0.01 & 42.56 ± 0.23 \\ 
STEVE\citep{singh2022simple}        & 25.37 ± 0.06 & 33.99 ± 0.01 & 36.83 ± 0.01 & 45.76 ± 0.06 \\ 
Slot-Diffusion\citep{wu2023slotdiffusion} & 30.67 ± 0.38 & 40.57 ± 0.15 & 43.96 ± 0.16 & 52.94 ± 0.22 \\ 
Video-Saur\citep{zadaianchuk2024object}   & 42.31 ± 0.09 & 53.04 ± 0.07 & 56.81 ± 0.32 & 68.68 ± 0.39 \\ 
Slot-BERT (ours) & \textbf{44.37 ± 0.27} & \textbf{55.31 ± 0.09} & \textbf{61.68 ± 0.25} & \textbf{73.22 ± 0.24} \\ \hline
\end{tabular}
}
\end{table}
To evaluate the robustness and generalization of our method to longer video sequences, we tested our model on sequences of 7 and 11 frames after training on 5 frames. A sliding window approach was employed to handle longer inputs, ensuring temporal alignment during prediction. Additionally, we introduced a future slot prediction mechanism that replaced the simple RNN-based slot initialization which is only possible  when testing on video length that is equal to length in training. In this new design, the temporal slot transformer is fed all previous $T-1$ slots, leaving the slot at the last position empty. The temporal slot transformer then predicts this missing slot, which serves as the initialization for subsequent frames. An illustration of this next slot initialization can be found in supplementary material section \ref{sec_supp_x_slot}.

The results in Table \ref{tab_longer} demonstrate that our method achieves better performance compared to state-of-the-art approaches across all metrics, particularly in terms of mBO-V, mBO-F, MBHD and FG-ARI. For instance, in 7-frame sequences, our method achieved an mBO-V of 48.0\% and an FG-ARI of 57.5\%, surpassing the next best method, Video-Saur, by 1.3\% and 1.4\%, respectively. When extended to 11-frame sequences, our method maintained its robust performance, achieving an mBO-V of 46.2\% and an FG-ARI of 56.8\%. These results highlight the minimal degradation in accuracy as the sequence length increases, demonstrating the adaptability of our approach.

With the future slot prediction mechanism, we observed further performance gains, particularly for longer sequences with 11 frames. The mBO-V increased to 46.9\%, and the FG-ARI improved to 57.3\%, highlighting the effectiveness of this advanced initialization strategy in maintaining temporal coherence.

Figure \ref{fig_longer} visualizes the impact of sequence length on accuracy. Across methods, mBO-F showed minimal variation with longer sequences, demonstrating that frame-level accuracy is generally stable across sequence lengths. However, the challenge of maintaining temporal consistency is evident in video-level metrics (mBO-V). Most methods, including SAVi, STEVE, and Video-Saur, exhibited a notable decline in mBO-V with longer sequences. In contrast, our method maintains the highest mBO-V among all evaluated methods. This robustness underscores our approach's ability to adapt to varying temporal contexts effectively.

{To further demonstrate performance on even longer temporal context windows, we train and test Slot-BERT on 30 second episodes. The results and comparison to other methods are shown in Table \ref{tab_30s}. Note that, compared to shorter videos (e.g., 11 frames), the video-level mBO score decreased by around 2\%. Nevertheless, Slot-BERT beats state-of-the-art approaches across all metrics and maintains a high image-level segmentation accuracy in comparison to training on shorter clips. Note that training on longer sequences is not necessarily beneficial once the temporal structure becomes less informative about frame-to-frame transitions. An optimal training protocol for Slot-BERT would thus take into account the length and complexity of typical surgical actions and tasks.}

\DIFadd{We evaluate long-range temporal tracking under more challenging conditions,
using the additional test set of unseen Cholecystectomy
videos, where frequent object occlusions as well as instrument entry and exit
occur. This setting reflects realistic surgical workflows and poses a
significant challenge for online object-centric tracking.
Long-range tracking performance is assessed using identity-aware metrics,
including IDF1 and Temporal Identity Persistence (T-IDP), together with
object localization accuracy (CorLoc).

As shown in Fig.~\ref{fig_tracking}, we evaluate tracking over sequences of 30 frames and
compare our full model against a baseline using vanilla RNN-based inference
without TST or contrastive learning. Our real-time method
improves average slot-level IDF1 by 9.4\%, T-IDP by 8.9\%, and CorLoc by 8.8\%,
demonstrating substantially stronger identity consistency and localization
accuracy under frequent object entry and exit.  }

\subsection{Ablation study  }

\DIFadd{
We conduct ablation studying by isolating the effects of our two main contributions, namely the slot contrastive objective and the TST module, on video segmentation accuracy and temporal tracking consistency across different datasets and video lengths.
} 

\DIFadd{
All ablation experiments are repeated across five independent runs,
resulting in 500-1000 total video-level evaluations per experiment.
Statistical significance is assessed using one-tailed Welch’s $t$-tests and Bonferroni correction for multiple comparisons. The observed improvements between the full model and
single-component ablations correspond to large effect sizes
(Cohen’s $d>0.8$), indicating that the gains are practically meaningful.
}

\begin{figure*}[t!]
    \centerline{
    \includegraphics[width=0.99\linewidth]{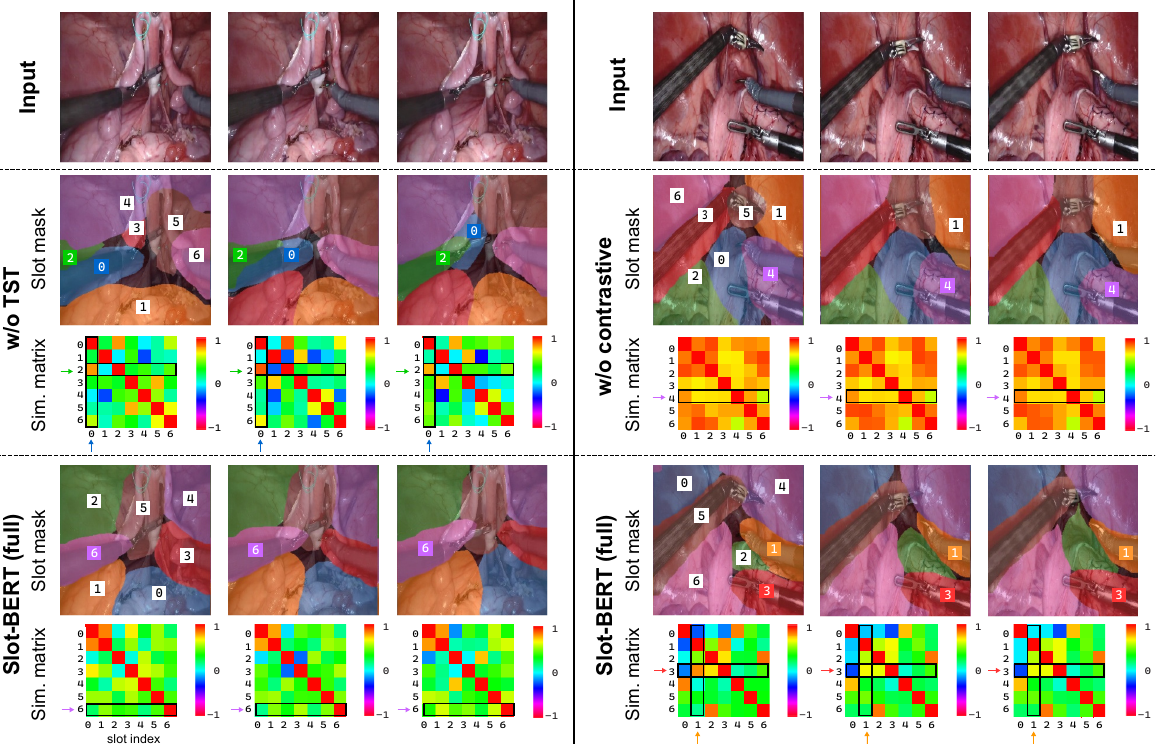}}
    \caption{{Slot masks and the corresponding cosine similarity matrices for Slot-BERT and ablated Slot-BERT variants without the TST module (left) or contrastive loss (right). All models are trained with 7 slots, and each segmentation mask is annotated with its corresponding slot index (0-6). See text for details.} } 
    \label{fig_sim_matrix}
\end{figure*}

{To illustrate the effects of the TST module and the contrastive loss, we compute the cosine similarity between slot representation vectors and visualize them as heat maps across frame sequences (Figure~\ref{fig_sim_matrix}). When the TST module is disabled, temporal consistency deteriorates. As shown on the left side of the figure, the mask originally tracking the instrument on the left (slot index 0, blue mask) switches to another slot (index 2, green mask). Similarly, the tissue at the bottom, initially represented by a single slot (index 1), is later fragmented into two slots (indices 1 and 3). These inconsistencies are also evident in the similarity matrix, where pronounced fluctuations appear in the rows corresponding to these slots. In contrast, under the full Slot-BERT configuration, object tracking is more stable. For example, the same left instrument (slot index 6, purple mask) remains consistently represented by a single slot across the entire sequence, which is reflected by the smoother, more stable row in the similarity matrix at the bottom.}

{The effect of removing the contrastive loss is shown on the right of Figure~\ref{fig_sim_matrix}. In this setting, slot similarities remain uniformly high (above 0.6), whereas with contrastive loss they are pushed lower, clustering around zero, enforcing that slots encode diverging directions in the representation hyper-space. Without contrastive loss, the model exhibits segmentation instability: two distinct instruments in the lower-right region of the frame are merged into a single slot (index 4) across the video. This ambiguity is reflected in the similarity matrix as bright regions indicating high correlation between slot vectors. By contrast, when contrastive loss is applied, slots are explicitly encouraged to diverge, enabling the two neighboring instruments to be represented by separate slots (indices 1 and 3, shown in orange and red masks).}

\begin{figure}[t!]
    \centerline{
    \includegraphics[width=0.9\linewidth]{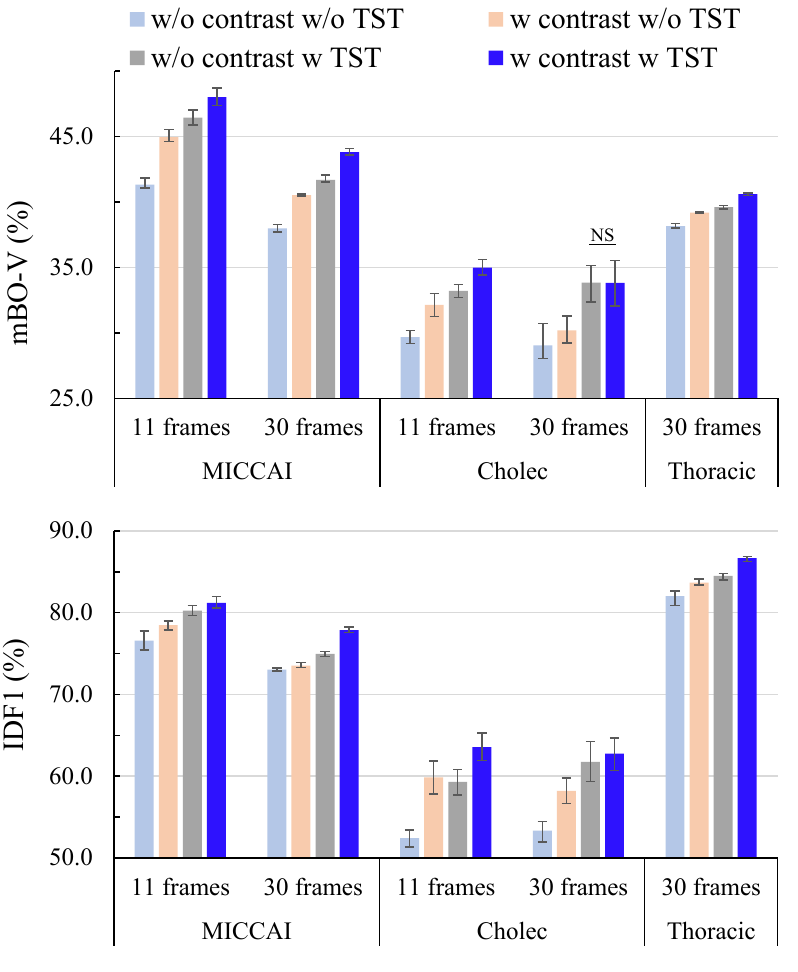}}
    \caption{ \DIFadd{Ablation study on three datasets with varying test video lengths. We compare video-level segmentation accuracy (mBO-V) and the temporal slot IDF1 score. Our full method is evaluated against variants with (w) or without (w/o) the slot contrastive learning objective and the TST module. Note that the Thoracic dataset is only partially and sparsely annotated; therefore, comparisons on this dataset are conducted only using the full video length (30 frames).} } 
    \label{fig_ablation_3data}
\end{figure}

\DIFadd{Figure~\ref{fig_ablation_3data} presents ablation results on three datasets with varying test-video lengths. We evaluate video-level accuracy (mBO-V) and temporal-slot IDF1, comparing the full method against variants with (w) or without (w/o) the contrastive objective and/or the TST module. Note that the Thoracic dataset is only partially and sparsely annotated; therefore, comparisons on this dataset are conducted only at full video length (30 frames).}

\DIFadd{
Across all datasets, both contrastive learning and temporal self-training
consistently improve surgical instrument segmentation and tracking.
On the MICCAI dataset (11-frame sequences), adding contrastive learning alone
improves mBO-V by approximately 3.6 percentage points over the baseline, while adding TST
alone yields an improvement of approximately 5.1 points; combining both
components results in a larger gain of approximately 6.7 points.
Similar trends are observed across datasets and metrics, with combined gains
of 3--7 points in mBO-V and 4--11 points in temporal slot IDF1.

Most of these improvements are statistically significant ($p < 0.0167$, with Bonferroni correction $m=3$).
An exception is observed for mBO-V on Cholec at 30 frames when comparing the
full model to the TST-only variant, where the difference does not reach
statistical significance ($p = 0.46$). This behavior likely reflects the increased
difficulty of segmenting tissue and small instruments in Cholec videos.
Importantly, even in this setting, the full model achieves significantly
higher temporal slot IDF1 than the TST-only variant ($p < 0.01$), indicating
stronger identity consistency.
Taken together, these results confirm the complementary roles of contrastive
learning and temporal self-training in robust surgical scene decomposition
and long-range tracking.}

\DIFdel{To investigate the contributions of individual components in our model to its final performance, we systematically removed or modified different components and observed their impact on segmentation accuracy, as summarized in Table 7. These experiments were conducted in both same-domain and zero-shot transfer scenarios, with evaluations on the MICCAI dataset (short and long sequences), EndoVis dataset, and Thoracic dataset. Below, we discuss the implications of the results for each configuration.}

\DIFdel{\textbf{Baseline:} }
\DIFdel{In this configuration, both the contrastive loss and \gls{tst} module were removed, and a simple recurrent refinement network (RRN) was employed along with a standard four-layer MLP decoder (denoted as w/o contrast w/o TST). This baseline yielded the lowest performance across all datasets and settings. For instance, on the MICCAI test set (long sequences), the mBO-V score dropped to $42.0\%$, while on EndoVis, it fell to $39.5\%$. These results highlight the critical contributions of both the contrastive loss and the \gls{tst} module in enhancing segmentation quality.}
 
\DIFdel{
\textbf{Slot Contrast only:} 
Removing the slot-BERT module while retaining the contrastive loss (denoted as contrast only) resulted in improvements over the vanilla baseline. For instance, on the MICCAI dataset (short sequences), the mBO-V score increased to $46.4\%$, and the FG-ARI improved to $55.3\%$. However, the performance remained below that of the full model, underscoring the role of slot-BERT in refining slot representations and contextualizing features effectively.

\textbf{Effect of slot contrastive loss:} 
In this configuration, the contrastive loss was omitted while retaining the slot-BERT module (denoted as w/o contrast). The results were similar to those of the previous configuration, with slight decreases in performance on most datasets. For example, on the EndoVis dataset, the mBO-V score was $40.8\%$, which is $1.3\%$ lower than the configuration without TST but with contrastive loss. These findings indicate that the contrastive loss is pivotal for improving slot distinctiveness, particularly in datasets with diverse instrument appearances.
}

\DIFdel{\textbf{Effect of slot masks:} 
We also examined the impact of slot-specific masking by directly inputting video slots into the \gls{tst} module without applying slot masks and replacing the masked transformer with a standard transformer (referred to as w/o slot masks). This led to a slight performance degradation. For instance, on the MICCAI dataset (long sequences), the mBO-V score was $45.9\%$ compared to $46.9\%$ for the full model. This suggests that slot masks do contribute to robust representation learning by focusing on relevant feature regions.

In addition, we replaced slot-specific masking with feature-level masking. A masked autoencoding strategy was applied, where random feature patches were masked during training (denoted as maskout feature). This configuration showed further degradation in performance compared to the full model. For example, on the Thoracic dataset, the mBO-V score was $32.9\%$, significantly lower than the $37.7\%$ achieved by the full model. This indicates that random feature masking is less effective than the structured masking strategy employed by the slot-BERT module.
 
 The results in Table 7 demonstrate the critical role of each component in our model. The full model consistently outperformed all other configurations across datasets and metrics. Notably, it achieved the highest mBO-V scores in all settings, including $48.9\%$ on MICCAI (short sequences), $44.0\%$ on EndoVis, and $37.7\%$ on Thoracic.  These results validate the synergistic contributions of the  TST module, contrastive loss, and slot-specific masking in achieving superior segmentation performance. Specifically, the contrastive loss enhances inter-slot separation, the \gls{tst} module improves feature contextualization and representation, and slot-specific masking ensures better focus on relevant regions. The full model's superior performance across both same-domain and zero-shot scenarios underscores the importance of integrating these components for robust segmentation in diverse domains.
}

\subsection{{Contrastive hyperparameter tuning analysis}}

\begin{figure}[t!]
    \centerline{
    \includegraphics[width=1.0\linewidth]{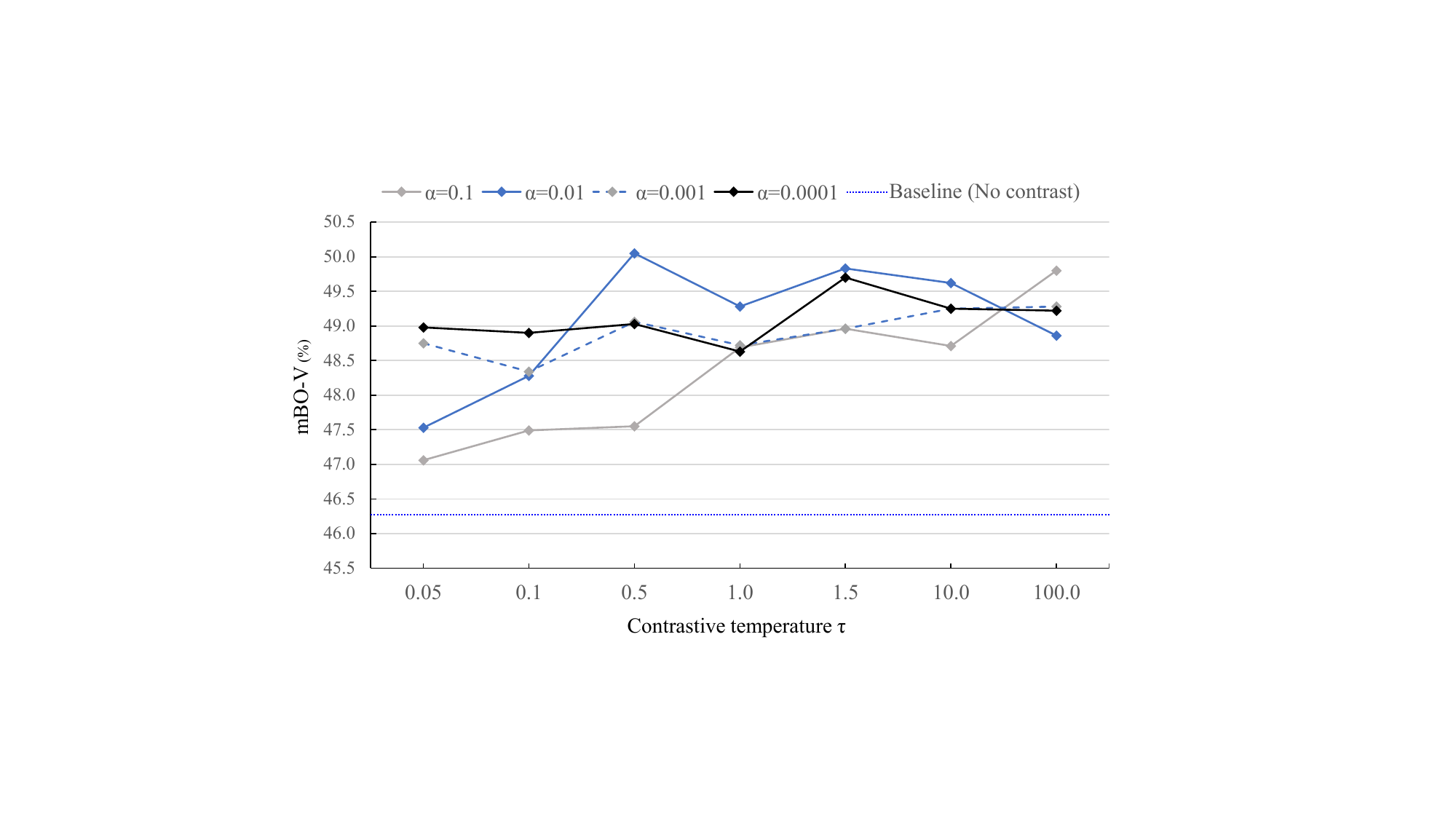}}
    \caption{ {Parameter tuning of Slot-BERT with respect to contrastive loss weight ($\alpha$) and temperature ($\tau$). Performance is reported as mBO-V. Across a range of $\alpha$ values (0.0001-0.1) and $\tau$ values (0.05-100), we observe consistent improvement compared to the baseline without contrastive learning. The best performance is achieved with $\alpha$ = 0.01 and $\tau$ = 0.5, reaching 50.1 mBO-V, highlighting the importance of jointly balancing temperature scaling and loss weighting. }} 
    \label{fig_tau_tune}
\end{figure}

{To investigate the effect of contrastive loss hyper-parameters on model performance, we systematically varied the contrastive loss weight ($\alpha$) from 0.0001 to 0.1 and the temperature parameter ($\tau$) from 0.05 to 100.  Figure~\ref{fig_tau_tune} visualizes the tendency.
}



{
Overall, we observe that a larger contrastive weight (e.g., $\alpha = 0.1$) requires a higher temperature to achieve better accuracy. This is likely because over-enforcing a strong dissimilarity between slots can have negative impact on object-centric learning; a higher temperature effectively smooths the contrastive distribution, mitigating the over-penalization caused by large weights. Conversely, when the contrastive weight is too small (e.g., $\alpha = 10^{-4}$), the effect of varying the temperature is marginal, as the relative change of influence on training dynamics is much less for this combination.

These findings suggest that a moderate weight on the contrastive objective, coupled with a balanced temperature scaling, provides the best trade-off between encouraging slot diversity and maintaining stable optimization. In particular, tuning the parameters to $\alpha = 0.01$ and $\tau = 0.5$ yields the highest video-level segmentation performance, achieving 50.1 mBO-V, which represents a notable improvement over the baseline model without contrastive learning (46.3 mBO-V).
}

\subsection{Experiment with Slot-Mixer decoder}
\label{result_mixer}
 \begin{table}[t!]
\caption{Performance comparison of Slot-Mixer decoder configurations on MICCAI and zero-shot Endovis benchmarks.
The table presents results for three configurations: Mixer Only, Contrast+Mixer, and Slot-BERT+Mixer. The Slot-BERT+Mixer configuration consistently outperforms the others across both benchmarks, demonstrating that for the alternative decoders the proposed slot contrastive learning with TST module can enhance object-centric learning performance in video.}
\label{tab_mixer}
\centering
\setlength\arrayrulewidth{0.9pt}
\setlength\doublerulesep{0.9pt} 
\resizebox{1.0\linewidth}{!}{%
\begin{tabular}{ccccc}
\hline
\multicolumn{1}{c}{\multirow{2}{*}{Setup}} & \multicolumn{4}{c}{MICCAI}                                                     \\ \cline{2-5} 
\multicolumn{1}{c}{}                       &mBO-V  & mBO-F    & mBHD($\downarrow$)           & CorLoc   \\ \hline
Mixer only                                 &47.1±0.4    & 51.2±0.3    & 50.889±0.326 & 61.6±0.5    \\
Contrast+Mixer                           &48.9±0.6    & 53.1±0.5    & 48.121±1.197 & 66.2±0.7    \\
 Mixer-full                         &\textbf{49.0±0.4}    & \textbf{53.2±0.2}    & \textbf{46.994±1.008} & \textbf{67.4±0.9}    \\ \hline
\multicolumn{1}{c}{\multirow{2}{*}{Setup}} & \multicolumn{4}{c}{Zero shot to Endovis}                                       \\ \cline{2-5} 
\multicolumn{1}{c}{}                       &mBO-V      & mBO-F    & mBHD ($\downarrow$)         & CorLoc   \\ \hline
Mixer only                                 &42.7±0.3    & 47.1±0.2    & 54.321±0.6   & 58.1±0.6    \\
Contrast+Mixer                           &\textbf{43.3±0.1}    &\textbf{47.5±0.2}   & 53.021±0.279 & 60.9±1.1    \\
 Mixer-full                         &43.2±0.2    &\textbf{47.5±0.2}    & \textbf{51.945±0.5}  & \textbf{61.9±0.3}    \\ \hline
\end{tabular}
}
\end{table}
To evaluate our method with the alternative Slot-Mixer decoder~\citep{zadaianchuk2024object}, we conducted experiments using the Slot-BERT framework and compared it against two additional configurations. The results are reported for the MICCAI benchmark, where the model is trained and tested on MICCAI, and the zero-shot Endovis benchmark, where the model trained on MICCAI is tested directly on Endovis without fine-tuning.
The first configuration, \textbf{Mixer Only}, combines a vanilla RNN Slot Attention with the Slot Mixer for feature reconstruction. The second, \textbf{Contrast+Mixer}, incorporates slot contrastive loss into the training process while still relying on the Slot-Mixer for decoding. Finally, \textbf{Mixer-full} represents the full proposed method.

As shown in table \ref{tab_mixer}, in the MICCAI benchmark, the proposed method achieves an mBHD error of 46.9, improving upon the results of baseline of Mixer-only  50.8. For Endovis, in zero-shot transfer, our method achieves mBHD of 51.9. This demonstrates the comparable performance of the Slot-Mixer decoder to the MLP decoder:44.2 mBHD on MICCAI and 50.7 mBHD zero shot to Endovis  (table \ref{tab_scratch} and table \ref{tab_zeroshot}).

These findings confirm that our method is able to surpass the vanilla RNN-based baseline using either MLP broadcast{,} or Slot-Mixer decoder {which improves decoding quality by introducing interactions between slots}, achieving high localization and segmentation performance. For more qualitative results refer to the section \ref{sec_quali_mlp_mix} of supplementary materials.

\subsection{Affect of the amount of training data on performance}

\begin{figure}[t!]
    \centerline{
    \includegraphics[width=1.0\linewidth]{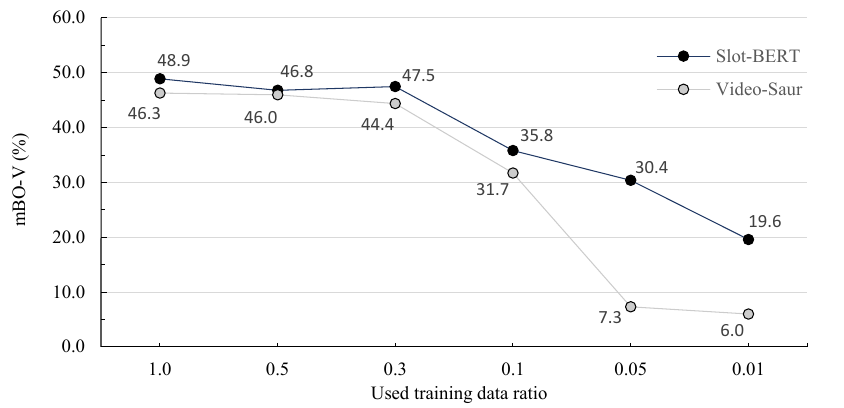}}
    \caption{Comparison of Slot-BERT and Video-Saur performance as a function of training data proportion. The evaluation metrics is mBO-V (video-level segmentation accuracy), with data proportions ranging from the full dataset (1.0) to 1\% (0.01). Slot-BERT demonstrates much higher accuracy (20.8 vs 6.1 mBO-V) even with minimal training data, whereas Video-Saur's performance significantly drops as data decreases. } 
    \label{fig_data_percent}
\end{figure}

We evaluate the sensitivity of segmentation performance to the amount of training data. Figure~\ref{fig_data_percent} compares Slot-BERT, our proposed method, with Video-Saur on the MICCAI challenge dataset, using mBO-V as video-level segmentation accuracy evaluation metrics. Training data proportions range from the full dataset (1.0) to 1\% (0.01).  

Slot-BERT demonstrates remarkable robustness in dealing with a decrease in training data. With the full dataset, it achieves 48.9\% accuracy compared to 46.3\% for Video-Saur. With only 30\% of the data, Slot-BERT maintains 47.5\%, nearly matching its full training data performance, while Video-Saur drops to 44.4\%. At lower data ratios, the gap widens significantly: Slot-BERT achieves 30.4\% at 0.05 compared to Video-Saur’s 7.3\%, and at 0.01, Slot-BERT retains 19.6\% while Video-Saur falls to 6.0\%.  
 In summary, Slot-BERT consistently outperforms Video-Saur across all data proportions and maintains higher accuracy even with minimal data. Its graceful degradation in performance highlights its robustness and makes it particularly effective for segmentation tasks in data-scarce scenarios.

\subsection{Experiment with different slot numbers}
\label{result_slot_num}
\begin{figure}[t!]
    \centerline{
    \includegraphics[width=1.0\linewidth]{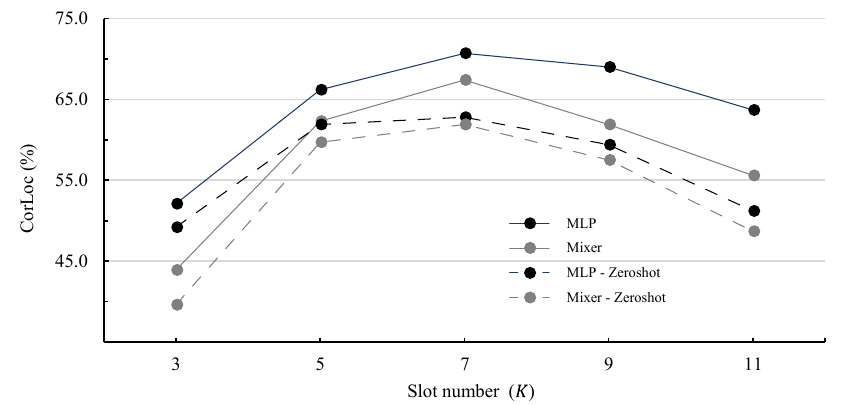}}
    \caption{Object localization accuracy (CorLoc) with varying slot numbers ($K$).
This figure presents the effect of changing the number of slots ($K$) on object localization accuracy, comparing performance on both the training domain and zero-shot data. The results show that $K = 7$ yields the best performance for both the MLP and Mixer decoders. } 
    \label{fig_slot_num}
\end{figure}
In order to investigate the effect of changing slot number $K$ on the performance of \gls{sbert}, experiments were conducted using MLP and Mixer decoders. Figure \ref{fig_slot_num} presents the CorLoc score obtained on training domain and zero-shot data, to evaluate impact of slot allocation on the localization accuracy. The results show that $K = 7$ generally provides the best performance for both methods, with MLP achieving a peak score of  70.7 and Mixer reaching 67.4 on training domain data. As $K$ increases beyond 7, performance begins to decline for both decoders, suggesting that excessively large $K$ values might over-delineate the image as the number of objects in a video is limited. MLP consistently outperforms Mixer on training domain data, indicating its superior ability to leverage increased $K$ for improved predictions.

On zero-shot data, the trend remains consistent, with both MLP and Mixer achieving their best scores at $K = 7$, though their overall performance is lower compared to the training domain. MLP slightly outperforms Mixer across most $K$values, reaching a maximum zero-shot score of 62.8, compared to Mixer's 61.9 at $K = 7$. These results highlight that while $K = 7$ is optimal across both datasets, there is a clear performance gap between training domain and zeroshot data, reflecting challenges in generalization. The analysis demonstrates that adjusting $K$ significantly impacts the effectiveness of both decoding methods, emphasizing the importance of finding an optimal $K$value to balance performance and generalization. Evaluation of different slot number under additional metrics are concluded in Supplementary Material section \ref{sec_supple_slot_num}.

\subsection{{Computational efficiency analysis}}
{We benchmark the computational efficiency of Slot-BERT against existing object-centric video models. The forward inference time per frame is measured on a single NVIDIA RTX A6000 GPU using video clips of $30$ seconds, where we set the temporal context window to $5$ seconds. We also report the floating point operations per second (GFLOPs), which quantify the computational cost of processing a single frame. The slot number for all methods is set to 7 in this experiment. 

As shown in Table~\ref{tab_runtime}, Video-Saur achieves the fastest inference speed ($1.2$ ms per frame), while Slot-BERT maintains a competitive runtime efficiency ($1.7$ ms per frame) with only a minor computational overhead compared to other baselines. Although Slot-Diffusion reports the lowest GFLOPs ($334.196$), its inference speed is not competitive due to the iterative nature of diffusion-based decoding process.}

{We also evaluated the memory consumption under real-time video processing with slot mixer decoder. In this setting, the RNN-based model, Video-Saur requires $369.32$ MB. By incorporating TST module With a 5-frame context window, the overhead of slot-BERT is only $1.69$ MB ($371.01$ MB total), and with a 30-frame context window, the overhead remains modest at $2.09$ MB, thanks to the latent space operation of temporal attention.}

\begin{table}[t!]
\caption{{Runtime and computational cost comparison in terms of per-frame inference time and GFLOPs. Best results are highlighted in bold.}}
\label{tab_runtime}
\centering
\setlength\arrayrulewidth{0.9pt}
\setlength\doublerulesep{0.9pt} 
\resizebox{1.0\linewidth}{!}{%
\begin{tabular}{lcc}
\hline
Method & Time per frame (ms) & GFLOPs \\ \hline
SAVi\citep{kipf2021conditional}           & 3.3 & 430.677 \\ 
STEVE\citep{singh2022simple}              & 8.2 & 540.701 \\ 
Slot-Diffusion\citep{wu2023slotdiffusion} & 5.7 & \textbf{334.196}\\ 
Video-Saur\citep{zadaianchuk2024object}   & \textbf{1.2} & 411.703 \\ 
Slot-BERT (ours)                          & 1.7 & 453.450 \\ \hline
\end{tabular}
}
\end{table}

\subsection{\DIFadd{Experiment with different context window lengths}}
\label{result_context}
\begin{figure}[t!]
    \centerline{
    \includegraphics[width=1.0\linewidth]{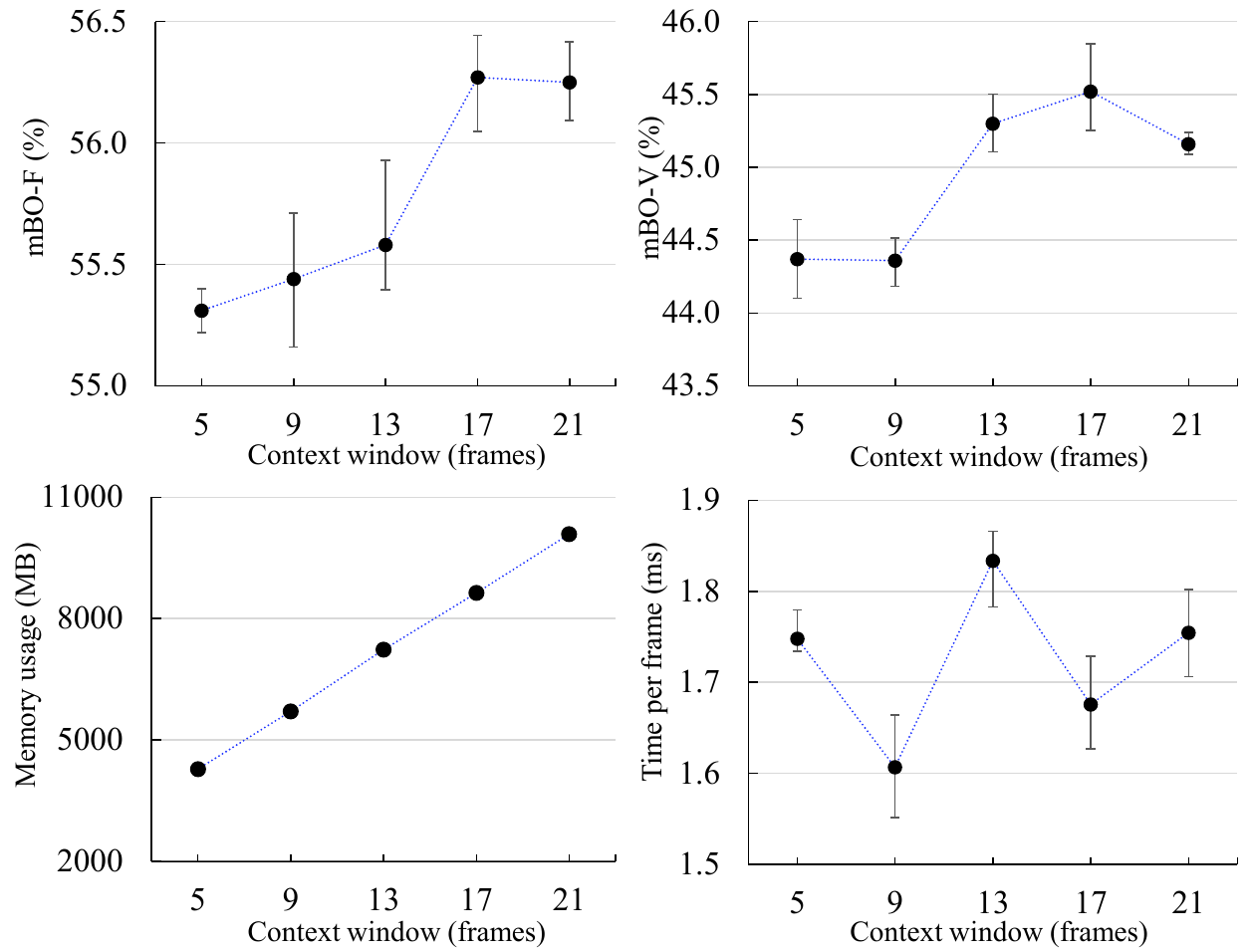}}
    \caption{ \DIFadd{Benchmark of our method using context window sizes ranging from 5 to 21 frames, evaluated on videos of fixed length (30 frames). Frame-level accuracy (mBO-F), video-level accuracy (mBO-V), GPU memory usage, and per-frame forward inference time are compared across different context window sizes.} } 
    \label{fig_context_len}
\end{figure}

\DIFadd{To analyze the trade-off between temporal context length, segmentation accuracy, and computational efficiency, we experiment with different context window lengths, ranging from 5 to 21 frames, while processing videos of fixed length (30 frames). As shown in Fig.~\ref{fig_context_len}, increasing the context window leads to up to 1\% improvements in both frame-level accuracy (mBO-F) and video-level accuracy (mBO-V), indicating that longer temporal context enhances reasoning over object dynamics.}

\DIFadd{
The performance gains are most pronounced when increasing the context window from 5 to 13 frames for the video-level metric mBO-V, which reflects a steady improvement in temporal consistency across frames. When the context window is further extended to 17 and 21 frames, the accuracy improvements begin to saturate, suggesting diminishing returns.

In contrast, the computational cost increases approximately linearly with the context window length, as reflected by the steady growth in GPU memory consumption. The per-frame inference time remains largely stable (around 1.7~ms) across different window lengths, indicating that the TST module introduces only minimal additional computational latency as the temporal context expands. The variations in per-frame runtime are thus driven by factors such as GPU parallelization and system-level scheduling rather than the choice of context window.}

\DIFadd{
Overall, this experiment highlights a clear accuracy-efficiency trade-off guiding the selection of context window length. On our datasets, a moderate context window (e.g., 9-13 frames) provides a favorable balance for unsupervised video segmentation, achieving most of the attainable accuracy gains while maintaining low memory overhead. At 1 FPS, the context window duration is between 9 to 13 seconds and this length seems to capture well the dy\-na\-mics of instruments and tissue in surgery. Ulti\-mately, the method is adaptable to different application scenarios and computational constraints.}
\section{Conclusions}
\label{sec:conclude}
 
\gls{sbert} demonstrated robust spatial segmentation and temporal reasoning capabilities, achieving state-of-the-art results across diverse datasets. The proposed bidirectional slots transformer enabled reasoning over video sequences without the scalability issues typical of parallel methods, because the attention operates on lightweight slot embeddings. This capability is particularly relevant for domains requiring long-term temporal coherence, such as surgical video analysis.  
 
By enforcing orthogonality between slots, our contrastive loss improved the independence of latent representations, reducing leakage across slot boundaries. This enhancement translated into more precise segmentation maps and improved object discovery, even in zero-shot settings.  
 
Unlike existing models that either simply update slots recursively, lacking long-range reasoning, or process entire video features for grouping, which is computationally costly, \gls{sbert} scales efficiently for videos. It achieves competitive performance without relying on computationally intensive modalities like optical flow or depth maps. This characteristic makes \gls{sbert} a practical choice for real-world applications requiring affordable computational resources. 

\subsection{Limitations}  

The effectiveness of slot-based representations partially depends on the quality of initial slot assignments. Suboptimal initialization can impact downstream temporal reasoning and segmentation quality. The number of slots also plays a role in how well slots encompass particular object instances. The optimal number of slots, as well as dynamic slot allocation are part of current research \citep{fan2024adaptive}. 

Although effective in handling moderate temporal dynamics, scenarios involving rapid, non-linear object motion or occlusions may still present challenges for \gls{sbert}. \DIFadd{Object exit from and re-entry into the field of view remains a fundamental
challenge for fully unsupervised, online object-centric tracking.
When an object disappears, its corresponding slot may be reassigned to
background regions. Upon reappearance, there is no explicit supervision to
guarantee reassociation with the original slot, particularly under prolonged
occlusion or appearance change.
Addressing this limitation may require incorporating memory mechanisms or
weak supervision \mbox{\citep{nwoye2019weakly,liao2025disentangling}}}. 

While our model outperformed baselines in zero-shot generalization tasks, there is room for improvement in handling novel object classes or extreme domain shifts.  

While our model discovers the overall location and shape of objects, it fails to predict the exact pixel-level boundaries, lowering the semantic segmentation accuracy. This is partially due to patch-based processing and could be mitigated by increasing the resolution of the patch grid. The model trained on lower-resolution path features can serve as guidance for higher-resolution modalities, such as optical flow and image saliency maps, enabling accurate high-resolution segmentation in an unsupervised manner. With \gls{sbert} the stability of integrating these modalities will enhanced. 

\subsection{Future Directions}

Incorporating additional priors or augmentations for slot initialization, such as unsupervised identifiable slot attention mechanisms \citep{kori2024identifiable}, may further enhance segmentation and temporal reasoning capabilities. Slot attention is a fully un-supervised representation learning algorithm; however, it can also be implemented in weakly supervised learning or in a human-interactive manner to identify the class of each slot so that pseudo-segmentation masks can be generated. Specifically, with weak video class labels, set prediction frameworks can be adopted to the latent space of slots \citep{sun2021rethinking}. Alternatively, with pre-trained slot models, annotators only need to identify which slot needs to be tracked or which slot represents the object of interest. In such interactive implementations, \gls{sbert} can reduce the annotation burden in labeling tasks or directly serve as an attention aid in tasks like surgical tracking and planning.

While \gls{sbert} eschews computationally intensive modalities like optical flow or feature similarity, integrating lightweight versions of these signals could provide complementary information in challenging scenarios. Such modalities could be an option to increase the resolution of slot attention masks, which are currently limited by the resolution of patches. For instance, the similarity of images or lower-level features often has much higher resolution compared to high-level features.

{Beyond incorporating additional modalities such as optical flow or depth maps to refine boundary accuracy, another promising direction is domain-specific high-resolution fine-tuning. This could be achieved by leveraging models pretrained on lower-resolution inputs (as typically required by default backbones like DINO) as anchors for adaptation~\mbox{\citep{didolkar2025transfer}}. In addition, DINOv3~\mbox{\citep{simeoni2025dinov3}} that is already trained with a high-resolution adaptation phase supports effective inference using high-resolution features.}

Extending \gls{sbert} to other biomedical domains, such as surgical videos involving cell segmentation in microscopic volumes or organ segmentation in MRI scanning sequences, would help evaluate its adaptability to diverse tasks.


\gls{sbert} represents a significant advancement in object-centric video representation learning, addressing critical challenges in long video sequences of surgical domains. By integrating slot-based reasoning with bidirectional transformers, the model balances scalability and segmentation accuracy, paving the way for future innovations in self-supervised learning. Its demonstrated potential for zero-shot generalization highlights the importance of modular, explainable architectures for tackling real-world, domain-specific challenges.

\section*{Acknowledgments}
This work was supported by the University of Pennsylvania Thomas B. McCabe and Jeannette E. Laws McCabe Fellow Award and the Linda Pechenik Montague Investigator Award.

\nocite{langley00}

\bibliography{bibliography/refs_rebib}

@inproceedings{ristani2016performance,
  title={Performance Measures and a Data Set for Multi-Target, Multi-Camera Tracking},
  author={Ristani, Ergys and Solera, Francesco and Zou, Roger and Cucchiara, Rita and Tomasi, Carlo},
  booktitle={European Conference on Computer Vision (ECCV)},
  year={2016}
}

@inproceedings{milan2016mot16,
  title={MOT16: A Benchmark for Multi-Object Tracking},
  author={Milan, Anton and Leal-Taix{\'e}, Laura and Reid, Ian and Roth, Stefan and Schindler, Konrad},
  booktitle={CVPR},
  year={2016}
}

@inproceedings{liao2025disentangling,
  title={Disentangling spatio-temporal knowledge for weakly supervised object detection and segmentation in surgical video},
  author={Liao, Guiqiu and Jogan, Matjaz and Koushik, Sai and Eaton, Eric and Hashimoto, Daniel A},
  booktitle={2025 IEEE/CVF Winter Conference on Applications of Computer Vision (WACV)},
  pages={8013--8023},
  year={2025},
  organization={IEEE}
}

@article{simeoni2025dinov3,
  title={DINOv3},
  author={Sim{\'e}oni, Oriane and Vo, Huy V and Seitzer, Maximilian and Baldassarre, Federico and Oquab, Maxime and Jose, Cijo and Khalidov, Vasil and Szafraniec, Marc and Yi, Seungeun and Ramamonjisoa, Micha{\"e}l and others},
  journal={arXiv preprint arXiv:2508.10104},
  year={2025}
}

@inproceedings{didolkar2025transfer,
  title     = {On the Transfer of Object-Centric Representation Learning},
  author    = {Didolkar, Aniket and Zadaianchuk, Andrii and Goyal, Anirudh and Mozer, Mike and Bengio, Yoshua and Martius, Georg and Seitzer, Maximilian},
  booktitle = {International Conference on Learning Representations (ICLR)},
  year      = {2025},
  url       = {https://arxiv.org/abs/2408.09162}
}

@inproceedings{xu2019youtubevis,
  title={YouTube-VIS: A Large-Scale Video Instance Segmentation Benchmark},
  author={Xu, Ning and Yang, Linjie and Fan, Yuchen and Yang, Dingcheng and Yue, Ding and Liang, Yuchen and Price, Brian and Cohen, Scott and Huang, Thomas},
  booktitle={Proceedings of the IEEE/CVF International Conference on Computer Vision (ICCV)},
  pages={2012--2021},
  year={2019}
}

@inproceedings{greff2022kubric,
  title={Kubric: A scalable dataset generator},
  author={Greff, Klaus and Minderer, Matthias and Papamakarios, George and Engelcke, Martin and Hennigan, Tom and Buhmann, Joachim M and Doucet, Arnaud and Murphy, Kevin and Arbel{\'a}ez, Pablo and Oliver, Nate and Beattie, Charlie and Gamper, Hannes and Malinowski, Mateusz and Gelly, Sylvain and Unther, Martin and Kipf, Thomas and Dosovitskiy, Alexey},
  booktitle={Proceedings of the IEEE/CVF Conference on Computer Vision and Pattern Recognition (CVPR)},
  pages={3749--3761},
  year={2022}
}

@article{tenenbaum2011grow,
 author = {Tenenbaum, Joshua B and Kemp, Charles and Griffiths, Thomas L and Goodman, Noah D},
 journal = {science},
 number = {6022},
 pages = {1279--1285},
 title = {How to grow a mind: Statistics, structure, and abstraction},
 volume = {331},
 year = {2011}
}

@article{kahneman1992reviewing,
 author = {Kahneman, Daniel and Treisman, Anne and Gibbs, Brian J},
 journal = {Cognitive psychology},
 number = {2},
 pages = {175--219},
 title = {The reviewing of object files: Object-specific integration of information},
 volume = {24},
 year = {1992}
}

@article{burgess2019monet,
 author = {Burgess, Christopher P and Matthey, Loic and Watters, Nicholas and Kabra, Rishabh and Higgins, Irina and Botvinick, Matt and Lerchner, Alexander},
 journal = {ArXiv preprint},
 title = {Monet: Unsupervised scene decomposition and representation},
 volume = {abs/1901.11390},
 year = {2019}
}

@inproceedings{greff2019iodine,
 author = {Klaus Greff and
Rapha{\"{e}}l Lopez Kaufman and
Rishabh Kabra and
Nick Watters and
Christopher Burgess and
Daniel Zoran and
Loic Matthey and
Matthew Botvinick and
Alexander Lerchner},
 booktitle = {Proceedings of the 36th International Conference on Machine Learning,
{ICML} 2019, 9-15 June 2019, Long Beach, California, {USA}},
 pages = {2424--2433},
 series = {Proceedings of Machine Learning Research},
 title = {Multi-Object Representation Learning with Iterative Variational Inference},
 volume = {97},
 year = {2019}
}

@inproceedings{locatello2020object,
 author = {Francesco Locatello and
Dirk Weissenborn and
Thomas Unterthiner and
Aravindh Mahendran and
Georg Heigold and
Jakob Uszkoreit and
Alexey Dosovitskiy and
Thomas Kipf},
 booktitle = {Advances in Neural Information Processing Systems 33: Annual Conference
on Neural Information Processing Systems 2020, NeurIPS 2020, December
6-12, 2020, virtual},
 title = {Object-Centric Learning with Slot Attention},
 year = {2020}
}

@article{greff2020binding,
 author = {Greff, Klaus and Van Steenkiste, Sjoerd and Schmidhuber, J{\"u}rgen},
 journal = {ArXiv preprint},
 title = {On the binding problem in artificial neural networks},
 volume = {abs/2012.05208},
 year = {2020}
}

@inproceedings{seitzer2022bridging,
 author = {Maximilian Seitzer and
Max Horn and
Andrii Zadaianchuk and
Dominik Zietlow and
Tianjun Xiao and
Carl{-}Johann Simon{-}Gabriel and
Tong He and
Zheng Zhang and
Bernhard Sch{\"{o}}lkopf and
Thomas Brox and
Francesco Locatello},
 booktitle = {The Eleventh International Conference on Learning Representations,
{ICLR} 2023, Kigali, Rwanda, May 1-5, 2023},
 title = {Bridging the Gap to Real-World Object-Centric Learning},
 year = {2023}
}

@inproceedings{fan2024adaptive,
 author = {Fan, Ke and Bai, Zechen and Xiao, Tianjun and He, Tong and Horn, Max and Fu, Yanwei and Locatello, Francesco and Zhang, Zheng},
 booktitle = {Proceedings of the IEEE/CVF Conference on Computer Vision and Pattern Recognition},
 pages = {23062--23071},
 title = {Adaptive slot attention: Object discovery with dynamic slot number},
 year = {2024}
}

@article{mansouri2023object,
 author = {Mansouri, Amin and Hartford, Jason and Zhang, Yan and Bengio, Yoshua},
 journal = {ArXiv preprint},
 title = {Object-centric architectures enable efficient causal representation learning},
 volume = {abs/2310.19054},
 year = {2023}
}

@inproceedings{jiang2023object,
 author = {Jindong Jiang and
Fei Deng and
Gautam Singh and
Sungjin Ahn},
 booktitle = {Advances in Neural Information Processing Systems 36: Annual Conference
on Neural Information Processing Systems 2023, NeurIPS 2023, New Orleans,
LA, USA, December 10 - 16, 2023},
 title = {Object-Centric Slot Diffusion},
 year = {2023}
}

@inproceedings{wu2023slotdiffusion,
 author = {Ziyi Wu and
Jingyu Hu and
Wuyue Lu and
Igor Gilitschenski and
Animesh Garg},
 booktitle = {Advances in Neural Information Processing Systems 36: Annual Conference
on Neural Information Processing Systems 2023, NeurIPS 2023, New Orleans,
LA, USA, December 10 - 16, 2023},
 title = {SlotDiffusion: Object-Centric Generative Modeling with Diffusion Models},
 year = {2023}
}

@article{weis2020unmasking,
 author = {Weis, Marissa A and Chitta, Kashyap and Sharma, Yash and Brendel, Wieland and Bethge, Matthias and Geiger, Andreas and Ecker, Alexander S},
 journal = {ArXiv preprint},
 title = {Unmasking the inductive biases of unsupervised object representations for video sequences},
 volume = {abs/2006.07034},
 year = {2020}
}

@inproceedings{kipf2021conditional,
 author = {Thomas Kipf and
Gamaleldin Fathy Elsayed and
Aravindh Mahendran and
Austin Stone and
Sara Sabour and
Georg Heigold and
Rico Jonschkowski and
Alexey Dosovitskiy and
Klaus Greff},
 booktitle = {The Tenth International Conference on Learning Representations, {ICLR}
2022, Virtual Event, April 25-29, 2022},
 title = {Conditional Object-Centric Learning from Video},
 year = {2022}
}

@inproceedings{aydemir2023self,
 author = {G{\"{o}}rkay Aydemir and
Weidi Xie and
Fatma G{\"{u}}ney},
 booktitle = {Advances in Neural Information Processing Systems 36: Annual Conference
on Neural Information Processing Systems 2023, NeurIPS 2023, New Orleans,
LA, USA, December 10 - 16, 2023},
 title = {Self-supervised Object-Centric Learning for Videos},
 year = {2023}
}

@inproceedings{zadaianchuk2024object,
 author = {Andrii Zadaianchuk and
Maximilian Seitzer and
Georg Martius},
 booktitle = {Advances in Neural Information Processing Systems 36: Annual Conference
on Neural Information Processing Systems 2023, NeurIPS 2023, New Orleans,
LA, USA, December 10 - 16, 2023},
 title = {Object-Centric Learning for Real-World Videos by Predicting Temporal
Feature Similarities},
 year = {2023}
}

@inproceedings{wu2022slotformer,
 author = {Ziyi Wu and
Nikita Dvornik and
Klaus Greff and
Thomas Kipf and
Animesh Garg},
 booktitle = {The Eleventh International Conference on Learning Representations,
{ICLR} 2023, Kigali, Rwanda, May 1-5, 2023},
 title = {SlotFormer: Unsupervised Visual Dynamics Simulation with Object-Centric
Models},
 year = {2023}
}

@inproceedings{biza2023invariant,
 author = {Ondrej Biza and
Sjoerd van Steenkiste and
Mehdi S. M. Sajjadi and
Gamaleldin Fathy Elsayed and
Aravindh Mahendran and
Thomas Kipf},
 booktitle = {International Conference on Machine Learning, {ICML} 2023, 23-29 July
2023, Honolulu, Hawaii, {USA}},
 pages = {2507--2527},
 series = {Proceedings of Machine Learning Research},
 title = {Invariant Slot Attention: Object Discovery with Slot-Centric Reference
Frames},
 volume = {202},
 year = {2023}
}

@inproceedings{lee2024guided,
 author = {Lee, Minhyeok and Cho, Suhwan and Lee, Dogyoon and Park, Chaewon and Lee, Jungho and Lee, Sangyoun},
 booktitle = {Proceedings of the IEEE/CVF Conference on Computer Vision and Pattern Recognition},
 pages = {3807--3816},
 title = {Guided Slot Attention for Unsupervised Video Object Segmentation},
 year = {2024}
}

@inproceedings{qian2023semantics,
 author = {Rui Qian and
Shuangrui Ding and
Xian Liu and
Dahua Lin},
 booktitle = {{IEEE/CVF} International Conference on Computer Vision, {ICCV} 2023,
Paris, France, October 1-6, 2023},
 pages = {16629--16641},
 title = {Semantics Meets Temporal Correspondence: Self-supervised Object-centric
Learning in Videos},
 year = {2023}
}

@article{nwoye2019weakly,
  title={Weakly supervised convolutional LSTM approach for tool tracking in laparoscopic videos},
  author={Nwoye, Chinedu Innocent and Mutter, Didier and Marescaux, Jacques and Padoy, Nicolas},
  journal={International journal of computer assisted radiology and surgery},
  volume={14},
  number={6},
  pages={1059--1067},
  year={2019},
  publisher={Springer}
}

@inproceedings{bao2022discovering,
 author = {Zhipeng Bao and
Pavel Tokmakov and
Allan Jabri and
Yu{-}Xiong Wang and
Adrien Gaidon and
Martial Hebert},
 booktitle = {{IEEE/CVF} Conference on Computer Vision and Pattern Recognition,
{CVPR} 2022, New Orleans, LA, USA, June 18-24, 2022},
 pages = {11779--11788},
 title = {Discovering Objects that Can Move},
 year = {2022}
}

@inproceedings{bao2023object,
 author = {Zhipeng Bao and
Pavel Tokmakov and
Yu{-}Xiong Wang and
Adrien Gaidon and
Martial Hebert},
 booktitle = {{IEEE/CVF} Conference on Computer Vision and Pattern Recognition,
{CVPR} 2023, Vancouver, BC, Canada, June 17-24, 2023},
 pages = {22972--22981},
 title = {Object Discovery from Motion-Guided Tokens},
 year = {2023}
}

@inproceedings{singh2022simple,
 author = {Gautam Singh and
Yi{-}Fu Wu and
Sungjin Ahn},
 booktitle = {Advances in Neural Information Processing Systems 35: Annual Conference
on Neural Information Processing Systems 2022, NeurIPS 2022, New Orleans,
LA, USA, November 28 - December 9, 2022},
 title = {Simple Unsupervised Object-Centric Learning for Complex and Naturalistic
Videos},
 year = {2022}
}

@inproceedings{singh2024parallelized,
 author = {Singh, Gautam and Wang, Yue and Yang, Jiawei and Ivanovic, Boris and Ahn, Sungjin and Pavone, Marco and Che, Tong},
 booktitle = {Forty-first International Conference on Machine Learning},
 title = {Parallelized Spatiotemporal Slot Binding for Videos},
 year = {2024}
}

@inproceedings{elsayed2022savi++,
 author = {Gamaleldin F. Elsayed and
Aravindh Mahendran and
Sjoerd van Steenkiste and
Klaus Greff and
Michael C. Mozer and
Thomas Kipf},
 booktitle = {Advances in Neural Information Processing Systems 35: Annual Conference
on Neural Information Processing Systems 2022, NeurIPS 2022, New Orleans,
LA, USA, November 28 - December 9, 2022},
 title = {SAVi++: Towards End-to-End Object-Centric Learning from Real-World
Videos},
 year = {2022}
}

@inproceedings{kenton2019bert,
 author = {Devlin, Jacob  and
Chang, Ming-Wei  and
Lee, Kenton  and
Toutanova, Kristina},
 booktitle = {Proceedings of the 2019 Conference of the North {A}merican Chapter of the Association for Computational Linguistics: Human Language Technologies, Volume 1 (Long and Short Papers)},
 pages = {4171--4186},
 title = {{BERT}: Pre-training of Deep Bidirectional Transformers for Language Understanding},
 year = {2019}
}

@inproceedings{kipf2019contrastive,
 author = {Thomas N. Kipf and
Elise van der Pol and
Max Welling},
 booktitle = {8th International Conference on Learning Representations, {ICLR} 2020,
Addis Ababa, Ethiopia, April 26-30, 2020},
 title = {Contrastive Learning of Structured World Models},
 year = {2020}
}

@inproceedings{sabour2017dynamic,
 author = {Sara Sabour and
Nicholas Frosst and
Geoffrey E. Hinton},
 booktitle = {Advances in Neural Information Processing Systems 30: Annual Conference
on Neural Information Processing Systems 2017, December 4-9, 2017,
Long Beach, CA, {USA}},
 pages = {3856--3866},
 title = {Dynamic Routing Between Capsules},
 year = {2017}
}

@inproceedings{hinton2018matrix,
 author = {Geoffrey E. Hinton and
Sara Sabour and
Nicholas Frosst},
 booktitle = {6th International Conference on Learning Representations, {ICLR} 2018,
Vancouver, BC, Canada, April 30 - May 3, 2018, Conference Track Proceedings},
 title = {Matrix capsules with {EM} routing},
 year = {2018}
}

@inproceedings{tsai2020capsules,
 author = {Yao{-}Hung Hubert Tsai and
Nitish Srivastava and
Hanlin Goh and
Ruslan Salakhutdinov},
 booktitle = {8th International Conference on Learning Representations, {ICLR} 2020,
Addis Ababa, Ethiopia, April 26-30, 2020},
 title = {Capsules with Inverted Dot-Product Attention Routing},
 year = {2020}
}

@inproceedings{henaff2022object,
 author = {H{\'e}naff, Olivier J and Koppula, Skanda and Shelhamer, Evan and Zoran, Daniel and Jaegle, Andrew and Zisserman, Andrew and Carreira, Jo{\~a}o and Arandjelovi{\'c}, Relja},
 booktitle = {European conference on computer vision},
 organization = {Springer},
 pages = {123--143},
 title = {Object discovery and representation networks},
 year = {2022}
}

@inproceedings{xu2022groupvit,
 author = {Xu, Jiarui and De Mello, Shalini and Liu, Sifei and Byeon, Wonmin and Breuel, Thomas and Kautz, Jan and Wang, Xiaolong},
 booktitle = {Proceedings of the IEEE/CVF Conference on Computer Vision and Pattern Recognition},
 pages = {18134--18144},
 title = {Groupvit: Semantic segmentation emerges from text supervision},
 year = {2022}
}

@inproceedings{kingma2013auto,
 author = {Diederik P. Kingma and
Max Welling},
 booktitle = {2nd International Conference on Learning Representations, {ICLR} 2014,
Banff, AB, Canada, April 14-16, 2014, Conference Track Proceedings},
 title = {Auto-Encoding Variational Bayes},
 year = {2014}
}

@inproceedings{mathieu2019disentangling,
 author = {Emile Mathieu and
Tom Rainforth and
N. Siddharth and
Yee Whye Teh},
 booktitle = {Proceedings of the 36th International Conference on Machine Learning,
{ICML} 2019, 9-15 June 2019, Long Beach, California, {USA}},
 pages = {4402--4412},
 series = {Proceedings of Machine Learning Research},
 title = {Disentangling Disentanglement in Variational Autoencoders},
 volume = {97},
 year = {2019}
}

@inproceedings{eastwood2018framework,
 author = {Cian Eastwood and
Christopher K. I. Williams},
 booktitle = {6th International Conference on Learning Representations, {ICLR} 2018,
Vancouver, BC, Canada, April 30 - May 3, 2018, Conference Track Proceedings},
 title = {A Framework for the Quantitative Evaluation of Disentangled Representations},
 year = {2018}
}

@inproceedings{kim2018disentangling,
 author = {Hyunjik Kim and
Andriy Mnih},
 booktitle = {Proceedings of the 35th International Conference on Machine Learning,
{ICML} 2018, Stockholmsm{\"{a}}ssan, Stockholm, Sweden, July 10-15,
2018},
 pages = {2654--2663},
 series = {Proceedings of Machine Learning Research},
 title = {Disentangling by Factorising},
 volume = {80},
 year = {2018}
}

@inproceedings{higgins2017beta,
 author = {Irina Higgins and
Lo{\"{\i}}c Matthey and
Arka Pal and
Christopher Burgess and
Xavier Glorot and
Matthew Botvinick and
Shakir Mohamed and
Alexander Lerchner},
 booktitle = {5th International Conference on Learning Representations, {ICLR} 2017,
Toulon, France, April 24-26, 2017, Conference Track Proceedings},
 title = {beta-VAE: Learning Basic Visual Concepts with a Constrained Variational
Framework},
 year = {2017}
}

@inproceedings{lin2020space,
 author = {Zhixuan Lin and
Yi{-}Fu Wu and
Skand Vishwanath Peri and
Weihao Sun and
Gautam Singh and
Fei Deng and
Jindong Jiang and
Sungjin Ahn},
 booktitle = {8th International Conference on Learning Representations, {ICLR} 2020,
Addis Ababa, Ethiopia, April 26-30, 2020},
 title = {{SPACE:} Unsupervised Object-Oriented Scene Representation via Spatial
Attention and Decomposition},
 year = {2020}
}

@article{van2020investigating,
 author = {Van Steenkiste, Sjoerd and Kurach, Karol and Schmidhuber, J{\"u}rgen and Gelly, Sylvain},
 journal = {Neural Networks},
 pages = {309--325},
 title = {Investigating object compositionality in generative adversarial networks},
 volume = {130},
 year = {2020}
}

@inproceedings{engelcke2019genesis,
 author = {Martin Engelcke and
Adam R. Kosiorek and
Oiwi Parker Jones and
Ingmar Posner},
 booktitle = {8th International Conference on Learning Representations, {ICLR} 2020,
Addis Ababa, Ethiopia, April 26-30, 2020},
 title = {{GENESIS:} Generative Scene Inference and Sampling with Object-Centric
Latent Representations},
 year = {2020}
}

@inproceedings{he2022masked,
 author = {Kaiming He and
Xinlei Chen and
Saining Xie and
Yanghao Li and
Piotr Doll{\'{a}}r and
Ross B. Girshick},
 booktitle = {{IEEE/CVF} Conference on Computer Vision and Pattern Recognition,
{CVPR} 2022, New Orleans, LA, USA, June 18-24, 2022},
 pages = {15979--15988},
 title = {Masked Autoencoders Are Scalable Vision Learners},
 year = {2022}
}

@inproceedings{caron2021emerging,
 author = {Mathilde Caron and
Hugo Touvron and
Ishan Misra and
Herv{\'{e}} J{\'{e}}gou and
Julien Mairal and
Piotr Bojanowski and
Armand Joulin},
 booktitle = {2021 {IEEE/CVF} International Conference on Computer Vision, {ICCV}
2021, Montreal, QC, Canada, October 10-17, 2021},
 pages = {9630--9640},
 title = {Emerging Properties in Self-Supervised Vision Transformers},
 year = {2021}
}

@article{ibrahim2023sat3d,
 author = {Ibrahim, Muhammad and Akhtar, Naveed and Anwar, Saeed and Mian, Ajmal},
 journal = {IEEE Transactions on Intelligent Transportation Systems},
 number = {5},
 pages = {5456--5466},
 title = {SAT3D: Slot attention transformer for 3D point cloud semantic segmentation},
 volume = {24},
 year = {2023}
}

@inproceedings{ponimatkin2023simple,
 author = {Ponimatkin, Georgy and Samet, Nermin and Xiao, Yang and Du, Yuming and Marlet, Renaud and Lepetit, Vincent},
 booktitle = {Proceedings of the IEEE/CVF Winter Conference on Applications of Computer Vision},
 pages = {5892--5903},
 title = {A simple and powerful global optimization for unsupervised video object segmentation},
 year = {2023}
}

@inproceedings{liu2021emergence,
 author = {Runtao Liu and
Zhirong Wu and
Stella X. Yu and
Stephen Lin},
 booktitle = {Advances in Neural Information Processing Systems 34: Annual Conference
on Neural Information Processing Systems 2021, NeurIPS 2021, December
6-14, 2021, virtual},
 pages = {13137--13152},
 title = {The Emergence of Objectness: Learning Zero-shot Segmentation from
Videos},
 year = {2021}
}

@inproceedings{karazija2022unsupervised,
 author = {Laurynas Karazija and
Subhabrata Choudhury and
Iro Laina and
Christian Rupprecht and
Andrea Vedaldi},
 booktitle = {Advances in Neural Information Processing Systems 35: Annual Conference
on Neural Information Processing Systems 2022, NeurIPS 2022, New Orleans,
LA, USA, November 28 - December 9, 2022},
 title = {Unsupervised Multi-Object Segmentation by Predicting Probable Motion
Patterns},
 year = {2022}
}

@inproceedings{choudhury2022guess,
 author = {Subhabrata Choudhury and
Laurynas Karazija and
Iro Laina and
Andrea Vedaldi and
Christian Rupprecht},
 booktitle = {33rd British Machine Vision Conference 2022, {BMVC} 2022, London,
UK, November 21-24, 2022},
 pages = {554},
 title = {Guess What Moves: Unsupervised Video and Image Segmentation by Anticipating
Motion},
 year = {2022}
}

@article{croitoru2019unsupervised,
 author = {Croitoru, Ioana and Bogolin, Simion-Vlad and Leordeanu, Marius},
 journal = {International Journal of Computer Vision},
 pages = {1279--1302},
 title = {Unsupervised learning of foreground object segmentation},
 volume = {127},
 year = {2019}
}

@inproceedings{uziel2023vit,
 author = {Roy Uziel and
Or Dinari and
Oren Freifeld},
 booktitle = {Advances in Neural Information Processing Systems 36: Annual Conference
on Neural Information Processing Systems 2023, NeurIPS 2023, New Orleans,
LA, USA, December 10 - 16, 2023},
 title = {From ViT Features to Training-free Video Object Segmentation via Streaming-data
Mixture Models},
 year = {2023}
}

@inproceedings{cheng2023tracking,
 author = {Ho Kei Cheng and
Seoung Wug Oh and
Brian Price and
Alexander G. Schwing and
Joon{-}Young Lee},
 booktitle = {{IEEE/CVF} International Conference on Computer Vision, {ICCV} 2023,
Paris, France, October 1-6, 2023},
 pages = {1316--1326},
 title = {Tracking Anything with Decoupled Video Segmentation},
 year = {2023}
}

@article{sestini2023fun,
 author = {Sestini, Luca and Rosa, Benoit and De Momi, Elena and Ferrigno, Giancarlo and Padoy, Nicolas},
 journal = {Medical Image Analysis},
 pages = {102751},
 title = {Fun-sis: A fully unsupervised approach for surgical instrument segmentation},
 volume = {85},
 year = {2023}
}

@inproceedings{vaswani2017attention,
 author = {Ashish Vaswani and
Noam Shazeer and
Niki Parmar and
Jakob Uszkoreit and
Llion Jones and
Aidan N. Gomez and
Lukasz Kaiser and
Illia Polosukhin},
 booktitle = {Advances in Neural Information Processing Systems 30: Annual Conference
on Neural Information Processing Systems 2017, December 4-9, 2017,
Long Beach, CA, {USA}},
 pages = {5998--6008},
 title = {Attention is All you Need},
 year = {2017}
}

@article{radford2018improving,
 author = {Radford, Alec},
 title = {Improving language understanding by generative pre-training},
 year = {2018}
}

@inproceedings{dosovitskiy2021image,
 author = {Alexey Dosovitskiy and
Lucas Beyer and
Alexander Kolesnikov and
Dirk Weissenborn and
Xiaohua Zhai and
Thomas Unterthiner and
Mostafa Dehghani and
Matthias Minderer and
Georg Heigold and
Sylvain Gelly and
Jakob Uszkoreit and
Neil Houlsby},
 booktitle = {9th International Conference on Learning Representations, {ICLR} 2021,
Virtual Event, Austria, May 3-7, 2021},
 title = {An Image is Worth 16x16 Words: Transformers for Image Recognition
at Scale},
 year = {2021}
}

@inproceedings{arnab2021vivit,
 author = {Anurag Arnab and
Mostafa Dehghani and
Georg Heigold and
Chen Sun and
Mario Lucic and
Cordelia Schmid},
 booktitle = {2021 {IEEE/CVF} International Conference on Computer Vision, {ICCV}
2021, Montreal, QC, Canada, October 10-17, 2021},
 pages = {6816--6826},
 title = {ViViT: {A} Video Vision Transformer},
 year = {2021}
}

@article{vincent2010stacked,
 author = {Vincent, Pascal and Larochelle, Hugo and Lajoie, Isabelle and Bengio, Yoshua and Manzagol, Pierre-Antoine and Bottou, Leon},
 journal = {Journal of Machine Learning Research},
 title = {Stacked denoising autoencoders: Learning useful representations in a deep network with a local denoising criterion},
 year = {2010}
}

@inproceedings{chen2020generative,
 author = {Mark Chen and
Alec Radford and
Rewon Child and
Jeffrey Wu and
Heewoo Jun and
David Luan and
Ilya Sutskever},
 booktitle = {Proceedings of the 37th International Conference on Machine Learning,
{ICML} 2020, 13-18 July 2020, Virtual Event},
 pages = {1691--1703},
 series = {Proceedings of Machine Learning Research},
 title = {Generative Pretraining From Pixels},
 volume = {119},
 year = {2020}
}

@inproceedings{bao2022beit,
 author = {Hangbo Bao and
Li Dong and
Songhao Piao and
Furu Wei},
 booktitle = {The Tenth International Conference on Learning Representations, {ICLR}
2022, Virtual Event, April 25-29, 2022},
 title = {BEiT: {BERT} Pre-Training of Image Transformers},
 year = {2022}
}

@inproceedings{dong2021peco,
 author = {Xiaoyi Dong and
Jianmin Bao and
Ting Zhang and
Dongdong Chen and
Weiming Zhang and
Lu Yuan and
Dong Chen and
Fang Wen and
Nenghai Yu and
Baining Guo},
 booktitle = {Thirty-Seventh {AAAI} Conference on Artificial Intelligence, {AAAI}
2023, Thirty-Fifth Conference on Innovative Applications of Artificial
Intelligence, {IAAI} 2023, Thirteenth Symposium on Educational Advances
in Artificial Intelligence, {EAAI} 2023, Washington, DC, USA, February
7-14, 2023},
 pages = {552--560},
 title = {PeCo: Perceptual Codebook for {BERT} Pre-training of Vision Transformers},
 year = {2023}
}

@inproceedings{wei2022masked,
 author = {Chen Wei and
Haoqi Fan and
Saining Xie and
Chao{-}Yuan Wu and
Alan L. Yuille and
Christoph Feichtenhofer},
 booktitle = {{IEEE/CVF} Conference on Computer Vision and Pattern Recognition,
{CVPR} 2022, New Orleans, LA, USA, June 18-24, 2022},
 pages = {14648--14658},
 title = {Masked Feature Prediction for Self-Supervised Visual Pre-Training},
 year = {2022}
}

@inproceedings{wang2022bevt,
 author = {Wang, Rui and Chen, Dongdong and Wu, Zuxuan and Chen, Yinpeng and Dai, Xiyang and Liu, Mengchen and Jiang, Yu-Gang and Zhou, Luowei and Yuan, Lu},
 booktitle = {Proceedings of the IEEE/CVF Conference on Computer Vision and Pattern Recognition},
 title = {BEVT: BERT pretraining of video transformers},
 year = {2022}
}

@inproceedings{ramesh2021zero,
 author = {Aditya Ramesh and
Mikhail Pavlov and
Gabriel Goh and
Scott Gray and
Chelsea Voss and
Alec Radford and
Mark Chen and
Ilya Sutskever},
 booktitle = {Proceedings of the 38th International Conference on Machine Learning,
{ICML} 2021, 18-24 July 2021, Virtual Event},
 pages = {8821--8831},
 series = {Proceedings of Machine Learning Research},
 title = {Zero-Shot Text-to-Image Generation},
 volume = {139},
 year = {2021}
}

@inproceedings{tong2022videomae,
 author = {Zhan Tong and
Yibing Song and
Jue Wang and
Limin Wang},
 booktitle = {Advances in Neural Information Processing Systems 35: Annual Conference
on Neural Information Processing Systems 2022, NeurIPS 2022, New Orleans,
LA, USA, November 28 - December 9, 2022},
 title = {VideoMAE: Masked Autoencoders are Data-Efficient Learners for Self-Supervised
Video Pre-Training},
 year = {2022}
}

@inproceedings{lee2019ficklenet,
 author = {Jungbeom Lee and
Eunji Kim and
Sungmin Lee and
Jangho Lee and
Sungroh Yoon},
 booktitle = {{IEEE} Conference on Computer Vision and Pattern Recognition, {CVPR}
2019, Long Beach, CA, USA, June 16-20, 2019},
 pages = {5267--5276},
 title = {FickleNet: Weakly and Semi-Supervised Semantic Image Segmentation
Using Stochastic Inference},
 year = {2019}
}

@inproceedings{hou2018self,
 author = {Qibin Hou and
Peng{-}Tao Jiang and
Yunchao Wei and
Ming{-}Ming Cheng},
 booktitle = {Advances in Neural Information Processing Systems 31: Annual Conference
on Neural Information Processing Systems 2018, NeurIPS 2018, December
3-8, 2018, Montr{\'{e}}al, Canada},
 pages = {547--557},
 title = {Self-Erasing Network for Integral Object Attention},
 year = {2018}
}

@inproceedings{fragkiadaki2015learning,
 author = {Katerina Fragkiadaki and
Pablo Arbelaez and
Panna Felsen and
Jitendra Malik},
 booktitle = {{IEEE} Conference on Computer Vision and Pattern Recognition, {CVPR}
2015, Boston, MA, USA, June 7-12, 2015},
 pages = {4083--4090},
 title = {Learning to segment moving objects in videos},
 year = {2015}
}

@inproceedings{tokmakov2017learning,
 author = {Pavel Tokmakov and
Karteek Alahari and
Cordelia Schmid},
 booktitle = {2017 {IEEE} Conference on Computer Vision and Pattern Recognition,
{CVPR} 2017, Honolulu, HI, USA, July 21-26, 2017},
 pages = {531--539},
 title = {Learning Motion Patterns in Videos},
 year = {2017}
}

@inproceedings{zhou2020motion,
 author = {Tianfei Zhou and
Shunzhou Wang and
Yi Zhou and
Yazhou Yao and
Jianwu Li and
Ling Shao},
 booktitle = {The Thirty-Fourth {AAAI} Conference on Artificial Intelligence, {AAAI}
2020, The Thirty-Second Innovative Applications of Artificial Intelligence
Conference, {IAAI} 2020, The Tenth {AAAI} Symposium on Educational
Advances in Artificial Intelligence, {EAAI} 2020, New York, NY, USA,
February 7-12, 2020},
 pages = {13066--13073},
 title = {Motion-Attentive Transition for Zero-Shot Video Object Segmentation},
 year = {2020}
}

@article{watters2019spatial,
 author = {Watters, Nicholas and Matthey, Loic and Burgess, Christopher P and Lerchner, Alexander},
 journal = {ArXiv preprint},
 title = {Spatial broadcast decoder: A simple architecture for learning disentangled representations in vaes},
 volume = {abs/1901.07017},
 year = {2019}
}

@inproceedings{sajjadi2022object,
 author = {Mehdi S. M. Sajjadi and
Daniel Duckworth and
Aravindh Mahendran and
Sjoerd van Steenkiste and
Filip Pavetic and
Mario Lucic and
Leonidas J. Guibas and
Klaus Greff and
Thomas Kipf},
 booktitle = {Advances in Neural Information Processing Systems 35: Annual Conference
on Neural Information Processing Systems 2022, NeurIPS 2022, New Orleans,
LA, USA, November 28 - December 9, 2022},
 title = {Object Scene Representation Transformer},
 year = {2022}
}

@article{zia2023surgical,
 author = {Zia, Aneeq and Bhattacharyya, Kiran and Liu, Xi and Berniker, Max and Wang, Ziheng and Nespolo, Rogerio and Kondo, Satoshi and Kasai, Satoshi and Hirasawa, Kousuke and Liu, Bo and others},
 journal = {ArXiv preprint},
 title = {Surgical tool classification and localization: results and methods from the MICCAI 2022 SurgToolLoc challenge},
 volume = {abs/2305.07152},
 year = {2023}
}

@article{twinanda2016endonet,
 author = {Twinanda, Andru P and Shehata, Sherif and Mutter, Didier and Marescaux, Jacques and De Mathelin, Michel and Padoy, Nicolas},
 journal = {IEEE transactions on medical imaging},
 number = {1},
 pages = {86--97},
 title = {Endonet: a deep architecture for recognition tasks on laparoscopic videos},
 volume = {36},
 year = {2016}
}

@article{hong2020cholecseg8k,
 author = {Hong, W-Y and Kao, C-L and Kuo, Y-H and Wang, J-R and Chang, W-L and Shih, C-S},
 journal = {ArXiv preprint},
 title = {Cholecseg8k: a semantic segmentation dataset for laparoscopic cholecystectomy based on cholec80},
 volume = {abs/2012.12453},
 year = {2020}
}

@article{liao2024disentangling,
 author = {Liao, Guiqiu and Jogan, Matjaz and Koushik, Sai and Eaton, Eric and Hashimoto, Daniel A},
 journal = {ArXiv preprint},
 title = {Disentangling spatio-temporal knowledge for weakly supervised object detection and segmentation in surgical video},
 volume = {abs/2407.15794},
 year = {2024}
}

@article{pont2016multiscale,
 author = {Pont-Tuset, Jordi and Arbelaez, Pablo and Barron, Jonathan T and Marques, Ferran and Malik, Jitendra},
 journal = {IEEE transactions on pattern analysis and machine intelligence},
 number = {1},
 pages = {128--140},
 title = {Multiscale combinatorial grouping for image segmentation and object proposal generation},
 volume = {39},
 year = {2016}
}

@inproceedings{rombach2022high,
 author = {Rombach, Robin and Blattmann, Andreas and Lorenz, Dominik and Esser, Patrick and Ommer, Bj{\"o}rn},
 booktitle = {Proceedings of the IEEE/CVF conference on computer vision and pattern recognition},
 pages = {10684--10695},
 title = {High-resolution image synthesis with latent diffusion models},
 year = {2022}
}

@article{van2008visualizing,
 author = {Van der Maaten, Laurens and Hinton, Geoffrey},
 journal = {Journal of machine learning research},
 number = {11},
 title = {Visualizing data using t-SNE.},
 volume = {9},
 year = {2008}
}

@article{kori2024identifiable,
 author = {Kori, Avinash and Locatello, Francesco and Santhirasekaram, Ainkaran and Toni, Francesca and Glocker, Ben and Ribeiro, Fabio De Sousa},
 journal = {Advances in Neural Information Processing Systems},
 title = {Identifiable Object-Centric Representation Learning via Probabilistic Slot Attention},
 year = {2024}
}

@inproceedings{xuslotVLM,
 author = {Xu, Jiaqi and Lan, Cuiling and Xie, Wenxuan and Chen, Xuejin and Lu, Yan},
 booktitle = {The Thirty-eighth Annual Conference on Neural Information Processing Systems},
 title = {Slot-VLM: Object-Event Slots for Video-Language Modeling},
 year = {2024}
}

@inproceedings{chen2020simple,
 author = {Ting Chen and
Simon Kornblith and
Mohammad Norouzi and
Geoffrey E. Hinton},
 booktitle = {Proceedings of the 37th International Conference on Machine Learning,
{ICML} 2020, 13-18 July 2020, Virtual Event},
 pages = {1597--1607},
 series = {Proceedings of Machine Learning Research},
 title = {A Simple Framework for Contrastive Learning of Visual Representations},
 volume = {119},
 year = {2020}
}

@inproceedings{bilen2016weakly,
 author = {Hakan Bilen and
Andrea Vedaldi},
 booktitle = {2016 {IEEE} Conference on Computer Vision and Pattern Recognition,
{CVPR} 2016, Las Vegas, NV, USA, June 27-30, 2016},
 pages = {2846--2854},
 title = {Weakly Supervised Deep Detection Networks},
 year = {2016}
}

@inproceedings{chen2021empirical,
 author = {Xinlei Chen and
Saining Xie and
Kaiming He},
 booktitle = {2021 {IEEE/CVF} International Conference on Computer Vision, {ICCV}
2021, Montreal, QC, Canada, October 10-17, 2021},
 pages = {9620--9629},
 title = {An Empirical Study of Training Self-Supervised Vision Transformers},
 year = {2021}
}

@inproceedings{assran2022masked,
 author = {Assran, Mahmoud and Caron, Mathilde and Misra, Ishan and Bojanowski, Piotr and Bordes, Florian and Vincent, Pascal and Joulin, Armand and Rabbat, Mike and Ballas, Nicolas},
 booktitle = {European Conference on Computer Vision},
 organization = {Springer},
 pages = {456--473},
 title = {Masked siamese networks for label-efficient learning},
 year = {2022}
}

@inproceedings{zhang2015self,
 author = {Dingwen Zhang and
Deyu Meng and
Chao Li and
Lu Jiang and
Qian Zhao and
Junwei Han},
 booktitle = {2015 {IEEE} International Conference on Computer Vision, {ICCV} 2015,
Santiago, Chile, December 7-13, 2015},
 pages = {594--602},
 title = {A Self-Paced Multiple-Instance Learning Framework for Co-Saliency
Detection},
 year = {2015}
}

@inproceedings{wang2017learning,
 author = {Lijun Wang and
Huchuan Lu and
Yifan Wang and
Mengyang Feng and
Dong Wang and
Baocai Yin and
Xiang Ruan},
 booktitle = {2017 {IEEE} Conference on Computer Vision and Pattern Recognition,
{CVPR} 2017, Honolulu, HI, USA, July 21-26, 2017},
 pages = {3796--3805},
 title = {Learning to Detect Salient Objects with Image-Level Supervision},
 year = {2017}
}

@inproceedings{sun2021rethinking,
 author = {Zhiqing Sun and
Shengcao Cao and
Yiming Yang and
Kris Kitani},
 booktitle = {2021 {IEEE/CVF} International Conference on Computer Vision, {ICCV}
2021, Montreal, QC, Canada, October 10-17, 2021},
 pages = {3591--3600},
 title = {Rethinking Transformer-based Set Prediction for Object Detection},
 year = {2021}
}

@article{lowe2022complex,
 author = {L{\"o}we, S and Lippe, P and Rudolph, M and Welling, M and others},
 journal = {Transactions on Machine Learning Research},
 number = {428},
 title = {Complex-Valued Autoencoders for Object Discovery},
 year = {2022}
}

@inproceedings{Kossen2020Structured,
 author = {Jannik Kossen and
Karl Stelzner and
Marcel Hussing and
Claas Voelcker and
Kristian Kersting},
 booktitle = {8th International Conference on Learning Representations, {ICLR} 2020,
Addis Ababa, Ethiopia, April 26-30, 2020},
 title = {Structured Object-Aware Physics Prediction for Video Modeling and
Planning},
 year = {2020}
}

@inproceedings{radford2021learning,
 author = {Alec Radford and
Jong Wook Kim and
Chris Hallacy and
Aditya Ramesh and
Gabriel Goh and
Sandhini Agarwal and
Girish Sastry and
Amanda Askell and
Pamela Mishkin and
Jack Clark and
Gretchen Krueger and
Ilya Sutskever},
 booktitle = {Proceedings of the 38th International Conference on Machine Learning,
{ICML} 2021, 18-24 July 2021, Virtual Event},
 pages = {8748--8763},
 series = {Proceedings of Machine Learning Research},
 title = {Learning Transferable Visual Models From Natural Language Supervision},
 volume = {139},
 year = {2021}
}
\bibliographystyle{template_double_column/icml2021}


\clearpage
\twocolumn[ 
    \begin{center}
        \Large \textbf{Supplementary Material}
    \end{center}
    \vspace{1em} 
]
\pagenumbering{arabic}

 
\renewcommand*{\thepage}{S\arabic{page}}
 \renewcommand\thefigure{S\arabic{figure}}  
 \renewcommand{\thetable}{S\arabic{table}}
\setcounter{figure}{0} 
\setcounter{table}{0} 

\setcounter{section}{0}
\renewcommand\thesection{\Alph{section}}
\renewcommand\thesubsection{\Alph{section}.\arabic{subsection}}
\newpage

\section{\DIFadd{Extended Implementation Details}}
\label{sec_supp_x_slot}
\DIFadd{
Section~\ref{sec:experiment_setup} presents the main experimental setup and model overview. Here, we provide additional implementation details covering network components, hyperparameters, and optimization settings across all experiments.
}
\paragraph{Network Architecture and Training Hyperparameters}
\DIFadd{
Our implementation is based on the Object Centric Learning framework with PyTorch. We set the same fixed learning rate ($1.0\times10^{-4}$) weight decay ($1.0\times10^{-5}$) and gradient clipping (0.05) for all modules for all the ablation studies and other parameter tuning (e.g. tuning of temperature and slot number parameter) experiments.  
}

\paragraph{Feature Encoder}  
\DIFadd{
We use a self-supervised Vision Transformer (ViT) as the feature encoder, following the work on object centric learning in real-world images or videos \mbox{\citep{seitzer2022bridging, zadaianchuk2024object}}. Specifically, we employ a DINO ViT backbone (\texttt{vit\_base\_patch16\_224\_dino})~\mbox{\citep{caron2021emerging}} that encodes $224 \times 224$ pixel resized images into patch embeddings. Patches are extracted with size $P = 16$ pixels, resulting in $N = 196$ spatial feature locations per frame.  
}

\paragraph{Slot Attention Module}
\DIFadd{
Slot Attention is applied independently per frame to decompose spatial features into $K$ latent object slots. Encoder features are linearly projected to the slot dimension and normalized. Slot Attention is iterated for $n_{\text{iters}}$ steps per frame using scaled dot-product attention, GRU-based slot updates, and an MLP with residual connections. First frame slots are randomly initialized for the first frame using a Gaussian distribution and propagated temporally for subsequent frames. Configuration details are summarized in Table~\ref{tab:slot_attention}.
}
 
\begin{table}[h]
\centering
\caption{\DIFadd{Slot Attention Module Configuration}}
\label{tab:slot_attention}
\setlength\arrayrulewidth{0.9pt}
\setlength\doublerulesep{0.9pt}
\resizebox{1.0\linewidth}{!}{%
\begin{tabular}{ll}
\hline
\textbf{Parameter} & \textbf{Value} \\
\hline
Number of slots ($K$) &   7 \\
Slot dimension ($d_{\text{slot}}$) & 64 \\
Slot attention iterations ($n_{\text{iters}}$) & 3 for first frame, 2 per other frames \\
Initialization strategy & Random Gaussian (first frame)  \\
Encoder output dim ($D_{\text{feat}}$) & 768 \\
MLP hidden dim &  [1024] \\
\hline
\end{tabular}}
\end{table}

\paragraph{\DIFadd{TST module}}
\DIFadd{
TST module operates on slot embeddings of shape $[B, T, K, d_{\text{slot}}]$. The TST consists of three transformer encoder layers with $8$ attention heads and feed-forward dimension $4\times d_{\text{slot}}$. Learnable temporal positional embeddings are added to slot representations. During training, random temporal masking with ratio $\gamma=0.20$ is applied for regularization. Full settings are listed in Table~\ref{tab:tst_module}.
}

 \begin{table}[h]
\centering
\caption{\DIFadd{Temporal Slot Transformer (TST) Module Configuration}}
\label{tab:tst_module}
\setlength\arrayrulewidth{0.9pt}
\setlength\doublerulesep{0.9pt}
\resizebox{1.0\linewidth}{!}{%
\begin{tabular}{ll}
\hline
\textbf{Parameter} & \textbf{Value} \\
\hline
Number of transformer layers & 3 \\
Attention heads per layer ($n_{\text{heads}}$) & 8 \\
Feed-forward hidden dimension & $4 \times d_{\text{slot}} = 256$ \\
Masking ratio ($\gamma$) & 0.20   \\
Temporal positional embeddings & Learnable, shape $[1, T, 1, d_{\text{slot}}]$ \\
Activation function & GELU \\
\hline
\end{tabular}}
\end{table}

\paragraph{\DIFadd{Slot Decoder}}
\DIFadd{
We consider an MLP decoder and a Slot-Mixer decoder. The MLP decoder independently maps each slot to spatial predictions using slot broadcasting and fully connected layers with additive positional embeddings. The Slot-Mixer decoder incorporates a transformer-based allocator (3 blocks, 4 attention heads) to jointly reason over slots prior to MLP-based rendering. Both decoders employ layer normalization for training stability. Unless otherwise stated, all parameter tuning and ablation studies are conducted using the MLP decoder. Detailed configurations are provided in Table~\ref{tab:slot_decoder}.
}
\begin{table}[h]
\centering
\caption{\DIFadd{Slot Decoder Variants Configuration}}
\label{tab:slot_decoder}
\setlength\arrayrulewidth{0.9pt}
\setlength\doublerulesep{0.9pt}
\resizebox{1.0\linewidth}{!}{%
\begin{tabular}{lll}
\hline
\textbf{Decoder Type} & \textbf{Component} & \textbf{Configuration} \\
\hline
\multirow{3}{*}{MLP}
 & Hidden layers & $[512, 512, 512, 512]$ \\
 & Additive position embedding & $[196, 64]$ \\
 & Normalization & Layer normalization \\
\hline
\multirow{4}{*}{Slot-Mixer}
 & Allocator & Transformer, 3 blocks, 4 heads \\
 & MLP renderer & $[1024, 1024]$, output dim 1024 \\
 & Additive position embedding & $[196, 64]$ \\
 & Normalization & Layer normalization \\
\hline
\end{tabular}}
\end{table}

\section{Illustration of using TST module for next slot initialization}
\label{sec_supp_x_slot}
\begin{figure}[h!]
    \centerline{
    \includegraphics[width=1.0\linewidth]{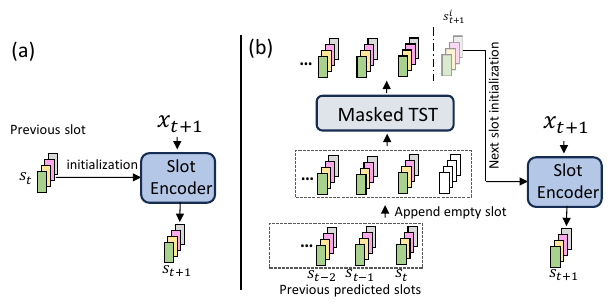}}
    \caption{Schematic illustration of next-slot initialization using TST. (a) Conventional slot initialization techniques, such as those used in RNN-based video slot attention algorithms. (b) Integration of same TST module as a slot initializer by appending an empty slot and masking the last embedding position. The predicted missing slot is then used to initialize the next slot prediction.} 
    \label{fig_x_slot}
\end{figure}

In Section \ref{sec_longer} of the main text, we present the results of adapting pre-trained slot-BERT for longer sequence prediction. In addition to using a traditional approach for slot initialization, where the previously predicted slot is fed into the next slot encoder as an initializer, we also demonstrate an alternative option of using the same \gls{tst} module as a slot initializer, as shown in Figure \ref{fig_x_slot}. In this new design, we take the previous slot buffer ${ ...,s_{t-1},s_t}$ and append an empty slot (zero vectors) to the latest position. The only modification to the TST module is to switch it from random masking to fixing the mask at the last location, enabling the TST module to predict the missing empty slot by reasoning over the historical slots. Finally, the next slot encoder takes the predicted slot from the TST module, $s^i_{t+1}$, as the initialization to update slot $s_{t+1}$. In our experiment, instead of creating a new TST module for the initializer or freezing the pre-trained TST for postprocessing, we allow the initializer and slot decoder to share weights from the same TST module and enable end-to-end training with the new slot initializer.

\begin{figure}[t!]
    \centerline{
    \includegraphics[width=1.0\linewidth]{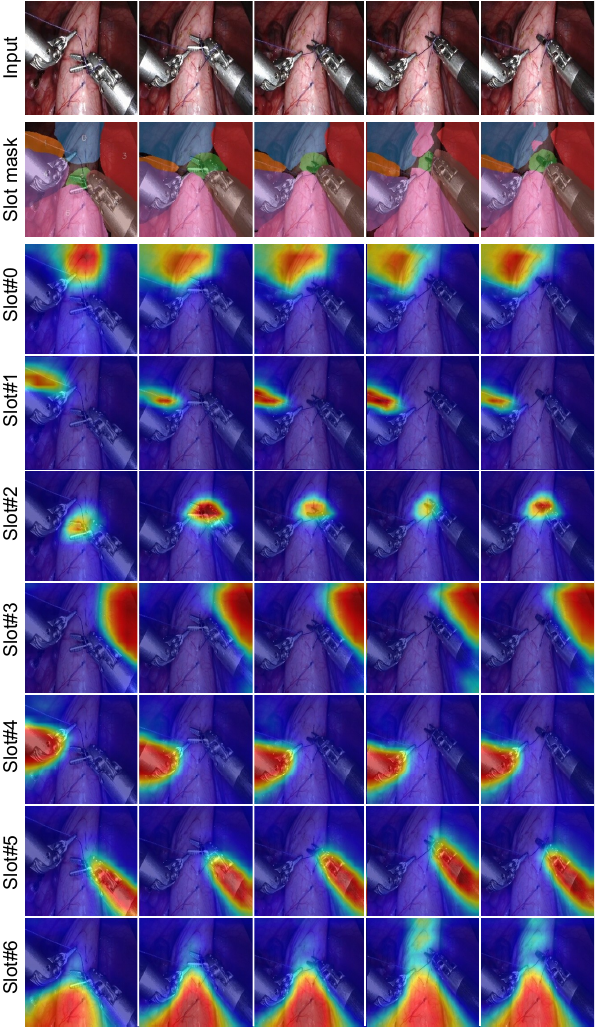}}
    \caption{{Visualization of multi-channel object-centric slot attention. We present both the binarized slot masks and the corresponding attention heatmaps for each slot. To indicate correspondence, the slot index is also overlaid in white on each colored mask of first frame.}} 
    \label{fig_attention}
\end{figure}
\section{ {Visualization of channel-wise object-centric attention overlay}}
{
We provide extended visualizations of the slot attention maps across independent channels (Figure \ref{fig_attention}). Following the presentation style of Figure~\ref{fig_sim_matrix}, to indicate correspondence, the slot index number is overlaid in white on each colored mask of the first frame. In this example, Slot-BERT is trained with seven slots for surgical videos, where these slots are allocated to attend to different instruments and tissue regions.
}

\section{ Visualization of latent slot embedding projection}

\begin{figure}[t!]
    \centerline{
    \includegraphics[width=1.0\linewidth]{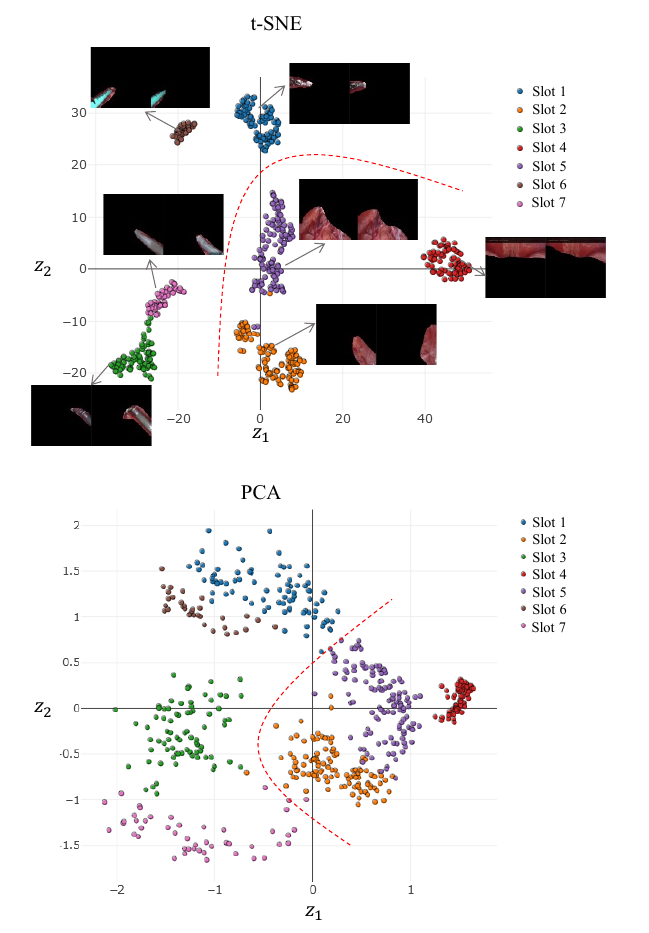}}
    \caption{Visualization of latent slot embeddings using t-SNE and PCA. Slot embeddings were extracted from a concatenated long video consisting of 100 frames (7 slots per frame) from the MICCAI challenge dataset. The scatter plots illustrate the clustering of slot vectors in the latent space. Example images decoded from corresponding slot masks are displayed alongside their projected points, demonstrating the separation of instrument slots and tissue slots into distinct and meaningful clusters.} 
    \label{fig_projection}
\end{figure}

To investigate the distinctiveness and relationships among slot vectors, we first visualize the latent slot embeddings learned by \gls{sbert} using t-SNE \citep{van2008visualizing}. As shown in Figure \ref{fig_projection}, slot vectors corresponding to different instruments within a video tend to cluster closer to each other, while remaining distinct from the clusters of tissue-related slot vectors. This behavior highlights the disentangled and interpretable nature of the slot representations learned by our approach.  

To further validate this observation, we applied Principal Component Analysis (PCA), a linear projection method that preserves the relative distances between data points in the latent space. The PCA mapping also reveals clear separation between regions corresponding to instrument slots and tissue slots, while maintaining the internal distinctiveness of each category. These results indicate that the slot embeddings learned by \gls{sbert} are both robust and explainable, effectively capturing the compositional structure of objects in video data.

\section{ Results of different slot numbers with additional metrics}
\label{sec_supple_slot_num}
 
\begin{table*}[t!]
\caption{This table presents the performance of our method with varying slot numbers ($K$) using both MLP and Slot-Mixer decoders across multiple evaluation metrics. The metrics include mBO-V, mBO-F, mBHD (boundary localization), FG-ARI, and CorLoc (spatial localization) in both the training and zero-shot domains. Notably, using an MLP decoder with 7 slots yields the highest CorLoc and mBHD scores in the training domain, while a  Mixer decoder with 9 slots shows higher mBO-F and mBO-V results in both domains.}
\label{tab_slot_num}
\centering
\setlength\arrayrulewidth{0.9pt}
\setlength\doublerulesep{0.9pt} 
\resizebox{1.0\linewidth}{!}{%
\begin{tabular}{ccccccccccccc}
\hline
\multicolumn{1}{l}{}        &             & \multicolumn{5}{c}{Training Domain}                                                                                     & \multicolumn{1}{c}{} & \multicolumn{5}{c}{Zero-Shot Domain}                                                                                   \\ \cline{3-7} \cline{9-13} 
\multicolumn{1}{l}{Decoder} & $K$         & mBO-V (\%)             & mBO-F (\%)             & mBHD ($\downarrow$)              & FG-ARI (\%)            & CorLoc (\%)            &                      & mBO-V (\%)             & mBO-F (\%)             & mBHD ($\downarrow$)              & FG-ARI (\%)            & CorLoc (\%)            \\ \hline
\multirow{5}{*}{MLP}        & 3           & 37.8 ± 0.3             & 40.5 ± 0.2             & 101.46 ± 0.93            & 42.3 ± 0.3             & 52.1 ± 0.4             &                      & 33.4 ± 0.2             & 35.7 ± 0.1             & 100.59 ± 0.33           & 38.8 ± 0.1             & 49.2 ± 0.5             \\
                            & 5           & 45.6 ± 0.4             & 48.9 ± 0.5             & 53.25 ± 2.21            & 53.1 ± 0.6             & 66.2 ± 0.8             &                      & 40.8 ± 0.1             & 44.2 ± 0.1             & 61.10 ± 0.35            & 49.7 ± 0.1             & 61.9 ± 0.5             \\
                            & 7           & 48.9 ± 0.2             & 52.8 ± 0.2             & 43.40 ± 0.53            & 58.2 ± 0.3             & \textbf{70.7 ± 0.8}    &                      & 44.0 ± 0.2             & 48.4 ± 0.1             & \textbf{49.26 ± 0.93}   & 55.2 ± 0.3             & \textbf{62.8 ± 0.2}    \\
                            & 9           & 49.1 ± 0.5             & \textbf{54.3 ± 0.1}    & \textbf{41.81 ± 0.94}   & \textbf{60.1 ± 0.2}    & 69.0 ± 0.5             &                      & 44.4 ± 0.1             & \textbf{49.4 ± 0.2}    & 49.43 ± 0.18            & 56.9 ± 0.3             & 59.4 ± 0.5             \\
                            & 11          & \textbf{49.2 ± 0.3}    & 53.6 ± 0.1             & 44.47 ± 0.19            & 60.0 ± 0.2             & 63.7 ± 0.8             &                      & \textbf{44.6 ± 0.4}    & 48.9 ± 0.1             & 50.82 ± 0.43            & \textbf{57.1 ± 0.2}    & 51.2 ± 0.5             \\ \hline
\multirow{5}{*}{Mixer}      & 3           & 33.9 ± 0.4             & 36.3 ± 0.3             & 114.97 ± 1.23           & 36.2 ± 0.4             & 43.9 ± 0.8             &                      & 30.0 ± 0.1             & 32.5 ± 0.1             & 110.21 ± 0.36           & 34.1 ± 0.1             & 39.6 ± 0.1             \\
                            & 5           & 45.2 ± 0.0             & 48.9 ± 0.1             & 59.85 ± 0.57            & 52.6 ± 0.2             & 62.3 ± 1.0             &                      & 40.2 ± 0.2             & 43.6 ± 0.2             & 63.13 ± 0.60            & 49.0 ± 0.1             & 59.7 ± 0.2             \\
                            & 7           & 49.0 ± 0.4             & 53.2 ± 0.2             & 46.99 ± 1.01            & 58.2 ± 0.2             & \textbf{67.4 ± 0.9}    &                      & 43.2 ± 0.2             & 47.5 ± 0.2             & 51.95 ± 0.50            & 54.3 ± 0.1             & \textbf{61.9 ± 0.3}    \\
                            & 9           & \textbf{49.3 ± 0.9}    & \textbf{53.7 ± 0.7}    & \textbf{46.41 ± 0.98}   & \textbf{59.7 ± 0.7}    & 61.9 ± 1.0             &                      & \textbf{44.0 ± 0.3}    & \textbf{49.1 ± 0.1}    & \textbf{51.03 ± 0.18}   & 56.6 ± 0.1             & 57.5 ± 0.5             \\
                            & 11          & 47.7 ± 0.6             & 52.5 ± 0.2             & 49.73 ± 0.62            & 59.1 ± 0.1             & 55.6 ± 0.6             &                      & 43.7 ± 0.3             & 48.8 ± 0.2             & 53.14 ± 0.65            & \textbf{57.1 ± 0.3}    & 48.7 ± 0.5             \\ \hline
\end{tabular}
}
\end{table*}

\begin{table*}[t!]
\caption{Transfer learning vs from scratch learning with Endovis and Thoracic dataset. The transfer learning uses models pre-trained on MICCAI dataset as initialization, and fine-tune them on the target domain data for 80 epoches. From scratch learning directly train the model on target domain for 2000 epoches.   }
\label{tab_tran2}
\centering
\setlength\arrayrulewidth{0.9pt}
\setlength\doublerulesep{0.9pt} 
\resizebox{0.9\linewidth}{!}{%
\begin{tabular}{clccccccccc}
\hline
\multicolumn{1}{l}{}      & \multicolumn{1}{l}{}                                 & \multicolumn{4}{c}{Transfer learning}                                                             &  & \multicolumn{4}{c}{From scratch}                                                                  \\ \cline{3-6} \cline{8-11} 
Dataset                   & Method                                               & \multicolumn{1}{c}{mBO-V(\%)} & \multicolumn{1}{c}{mBO-F(\%)} & \multicolumn{1}{c}{mBHD (↓)} & \multicolumn{1}{c}{FG-ARI(\%)} &  & \multicolumn{1}{c}{mBO-V(\%)} & \multicolumn{1}{c}{mBO-F(\%)} & \multicolumn{1}{c}{mBHD (↓)} & \multicolumn{1}{c}{FG-ARI(\%)} \\ \hline
\multirow{6}{*}{Endovis}  & DINO-Saur\citep{seitzer2022bridging}                 & 38.8                          & 44.1                          & 52.4                          & 51.3                           &  & 25.6                          & 26.1                          & 122.5                         & 29.4                           \\ 
                          & SAVi\citep{kipf2021conditional}                      & 30.3                          & 34.1                          & 77.8                          & 39.3                           &  & 34.4                          & 38.5                          & 110.9                         & 42.9                           \\ 
                          & STEVE\citep{singh2022simple}                         & 26.4                          & 30.3                          & 147.2                         & 34.5                           &  & 31.5                          & 34.1                          & 97.9                          & 38.6                           \\ 
                          & Slot-Diffusion\citep{wu2023slotdiffusion}            & 36.2                          & 38.5                          & 66.7                          & 43.3                           &  & 42.3                          & \textbf{46.0}                 & \textbf{56.9}                 & \textbf{53.2}                  \\ 
                          & Video-Saur\citep{zadaianchuk2024object}              & 46.9                          & 51.2                          & 50.8                          & 57.8                           &  & 25.7                          & 25.8                          & 135.0                         & 39.4                           \\ 
                          & Ours                                                 & \textbf{48.8}                 & \textbf{52.3}                 & \textbf{41.7}                 & \textbf{59.2}                  &  & \textbf{45.1}                 & \textbf{46.0}                 & 70.1                          & 51.6                           \\ \hline
\multirow{6}{*}{Thoracic} & DINO-Saur\citep{seitzer2022bridging}                 & 31.1                          & 39.3                          & 86.0                          & 32.5                           &  & 27.1                          & 30.7                          & 104.8                         & 23.9                           \\ 
                          & SAVi\citep{kipf2021conditional}                      & 25.6                          & 28.7                          & 106.5                         & 22.5                           &  & 26.5                          & 30.7                          & 119.8                         & 23.1                           \\ 
                          & STEVE\citep{singh2022simple}                         & 23.8                          & 30.0                          & 131.9                         & 22.3                           &  & 28.9                          & 33.0                          & 135.4                         & 26.5                           \\ 
                          & Slot-Diffusion\citep{wu2023slotdiffusion}            & 29.9                          & 37.8                          & 104.3                         & 28.8                           &  & 31.1                          & 39.9                          & \textbf{84.6}                 & 31.4                           \\ 
                          & Video-Saur\citep{zadaianchuk2024object}              & 38.9                          & \textbf{52.1}                 & 65.7                          & \textbf{41.9}                  &  & 21.9                          & 15.7                          & 139.5                         & 11.9                           \\ 
                          & Ours                                                 & \textbf{39.9}                 & \textbf{52.1}                 & \textbf{65.2}                 & 41.7                           &  & \textbf{34.0}                 & \textbf{40.5}                 & 92.4                          & \textbf{33.8}                  \\ \hline
\end{tabular}
}
\end{table*}

In section \ref{result_slot_num} we present the results of our method using different slot numbers. And demonstrated that with slot number 7 it shows optimal localization accuracy when either MLP or slot-Mixer decoder is adopted for \gls{sbert}. Here detailed quantitative results on additional metrics is revealed in table \ref{tab_slot_num}. As shown in Table \ref{tab_slot_num}, specifically, with 7 slots and an MLP decoder, the CorLoc metric reaches its peak at 70.7 ± 0.8 in the training domain, and mBHD achieves its value of 43.40 ± 0.53, reflecting accurate boundary localization. Similarly, with the Slot-Mixer decoder and 7 slots, the CorLoc metric achieves its highest value of 67.4 ± 0.9, demonstrating robust spatial localization.

When extending the analysis to evaluate segmentation overlap on the training and zero-shot domains, using 9 slots emerges as a strong performer. With the MLP decoder, 9 slots achieve the highest mBO-F at 54.3 ± 0.1 in the training domain and 49.4 ± 0.2 in the zero-shot domain,  Moreover, the Slot-Mixer decoder with 9 slots delivers the best mBO-V scores of 49.3 ± 0.9 for the training and zero-shot domains, respectively, signifying enhanced temporal coherence in video object segmentation.

These results underscore the nuanced trade-offs across different slot numbers and decoder configurations. While slot number 7 provides strong spatial localization, slot number 9 excels in overlap accuracy and temporal consistency, making it a versatile choice depending on the task emphasis.

\section{Additional results on transfer learning on Endovis and Thoracic data}

We fine-tuned the MICCAI pre-trained model on the EndoVis and Thoracic datasets for 80 epochs as an extention of the transfer learning experiment. Additionally, we trained all models from scratch on the EndoVis and Thoracic datasets for 2000 epochs, to compare their performance with transfer learning. To effectively utilize the small EndoVis and Thoracic datasets for fine-tuning, following previous unsupervised segmentation approaches \citep{zhang2015self, wang2017learning}, we allowed the model access to the full EndoVis and Thoracic videos. Our goal was to evaluate whether the models could generate accurate pseudo-masks under the condition of no ground truth supervision signals.

As presented in Table \ref{tab_tran2}, after fine-tuning on these two domain-specific datasets (EndoVis and Thoracic) using only self-supervised objectives such as reconstruction and contrastive learning, without relying on any labels, our method further improves performance. Remarkably, the fine-tuned results surpass those of models trained from scratch. For example, in EndoVis, our fine-tuned model achieves an FG-ARI of 59.2\%, which is higher than the scratch-trained model's FG-ARI of 51.6\%. Similarly, on the Thoracic dataset, our method achieves notable gains in mBO-V and FG-ARI after fine-tuning, demonstrating the effectiveness of leveraging knowledge obtained from unsupervised pretraining on MICCAI data.
It is worth noting that for some methods (e.g., Slot Diffusion and STEVE), transfer learning did not outperform training from scratch. This could possibly be because these methods tend to overfit on small datasets rather than generalize and effectively leverage knowledge across domains.

\begin{figure*}[t!]
    \centerline{
    \includegraphics[width=0.9\linewidth]{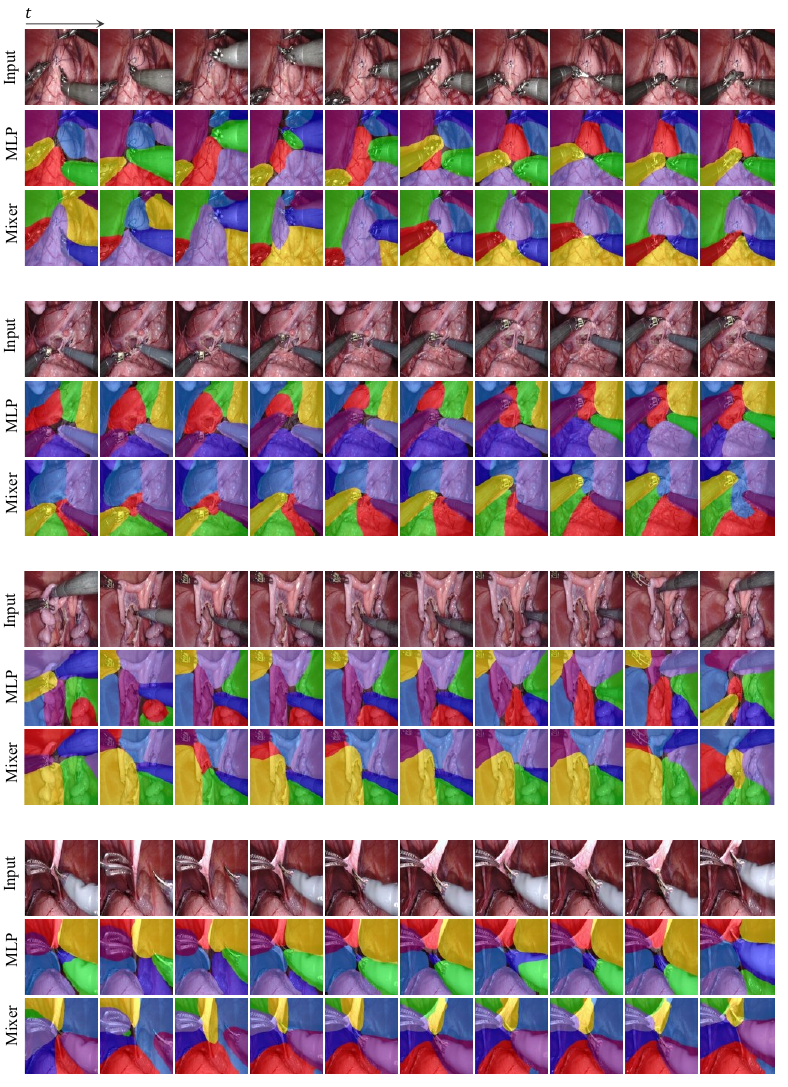}}
    \caption{Qualitative results of Slot-BERT with MLP and Slot-Mixerdecoder tested on longer sequences of 30 frames (1 FPS) from the MICCAI dataset. The model was trained with 7 slots. Each output sequence has been downsampled by a factor of 3, and 10 samples are displayed.} 
    \label{fig_miccaisupp}
\end{figure*}
\begin{figure*}[t!]
    \centerline{
    \includegraphics[width=0.9\linewidth]{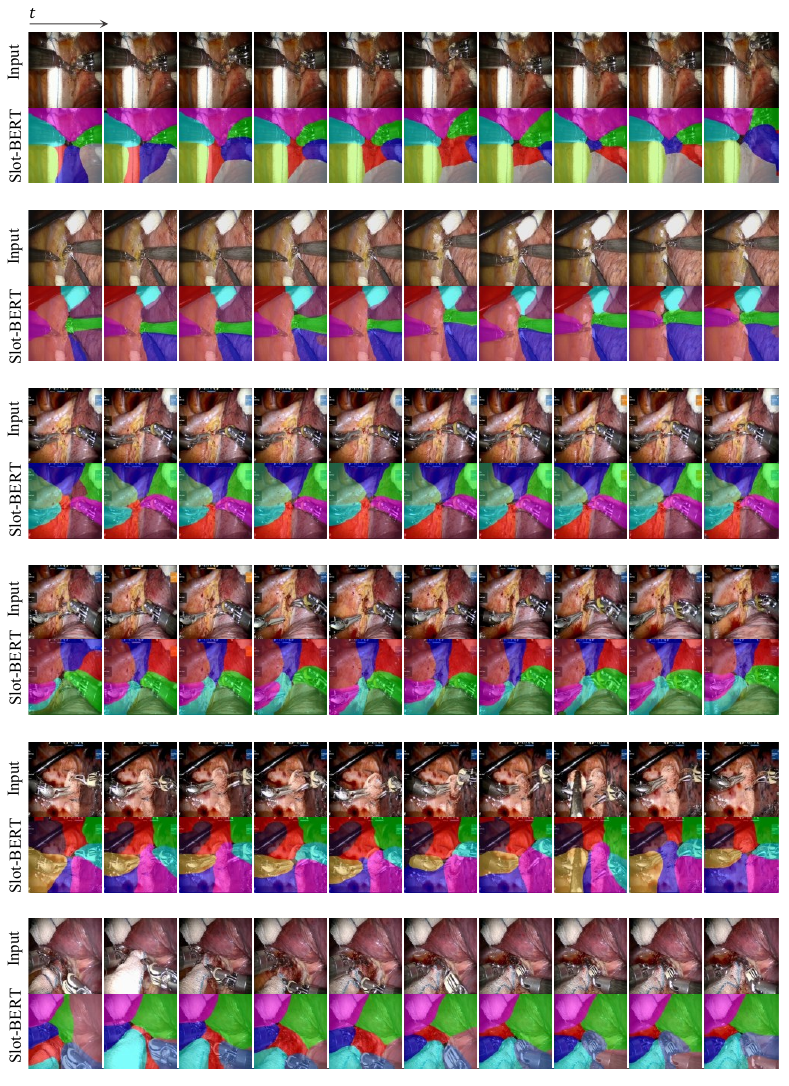}}
    \caption{Zero-shot transfer of Slot-BERT with Mixer decoder to thoracic robotic surgery videos after training on MICCAI abdominal data. The test sequence consists of 30 frames, downsampled to 10 frames for presentation. Notably, in some images, a slot successfully attends to gauze, an unseen object in the MICCAI dataset.} 
    \label{fig_0thora}
\end{figure*}

\begin{table*}[t!]
\caption{ \DIFadd{Additional ablation on the slot masks. The performance is tested under the training domain (MICCAI) and zero-shot transfer domain (EndoVis
and Thoracic) } }
\label{tab_add_ablation}
\centering
\setlength\arrayrulewidth{0.9pt}
\setlength\doublerulesep{0.9pt} 
\resizebox{1.0\linewidth}{!}{%
\begin{tabular}{@{}lccccccccccc@{}}
\toprule
                           & \multicolumn{5}{c}{Same domain (MICCAI)}                                                                    &           & \multicolumn{5}{c}{Zero shot transfer}                                                              \\ \cmidrule(lr){2-6} \cmidrule(l){8-12} 
                           & \multicolumn{2}{c}{ 5 frames }            &           & \multicolumn{2}{c}{ 11 frames }          &           & \multicolumn{2}{c}{Endovis}                    &  & \multicolumn{2}{c}{Thoracic}                    \\ \cmidrule(lr){2-3} \cmidrule(lr){5-6} \cmidrule(lr){8-9} \cmidrule(l){11-12} 
Method                     & mBO-V (\%)            & FG-ARI (\%)            &           & mBO-V (\%)           & FG-ARI (\%)            &           & mBO-V (\%)            & FG-ARI (\%)            &  & mBO-V (\%)             & FG-ARI (\%)            \\ \midrule
  
maskout feature            & 45.8 ± 0.5            & 55.0 ± 0.5             &           & 42.2 ± 0.6           & 53.6 ± 0.7             &           & 41.0 ± 0.5            & 51.7 ± 0.3             &  & 32.9 ± 0.2             & 36.6 ± 0.1             \\
w/o slot masks             & 48.6 ± 0.3            & 57.6 ± 0.5             &           & 45.9 ± 0.9           & 57.1 ± 0.4             &           & 43.0 ± 0.3            & 53.9 ± 0.2             &  & 35.0 ± 0.1             & 38.7 ± 0.1             \\
our full                   & \textbf{48.9 ± 0.2}   & \textbf{58.2 ± 0.3}    & \textbf{} & \textbf{46.9 ± 0.0}  & \textbf{57.3 ± 0.3}    & \textbf{} & \textbf{44.0 ± 0.2}   & \textbf{55.2 ± 0.3}    &  & \textbf{37.7 ± 0.2}    & \textbf{40.5 ± 0.1}    \\ \bottomrule
\end{tabular}
}
\end{table*}
\section{\DIFadd{Additional ablation study on slot masks}}
\DIFadd{
We also examined the impact of slot-specific masking by directly inputting video slots into the \gls{tst} module without applying slot masks and replacing the masked transformer with a standard transformer (referred to as w/o slot masks). This led to a slight performance degradation. As shown in Table \ref{tab_add_ablation}, on the MICCAI dataset (long sequences), the mBO-V score was $45.9\%$ compared to $46.9\%$ for the full model. This suggests that slot masks do contribute to robust representation learning by focusing on relevant feature regions.
}

\DIFadd{
In addition, we replaced slot-specific masking with feature-level masking. A masked autoencoding strategy was applied, where random feature patches were masked during training (denoted as maskout feature). This configuration showed further degradation in performance compared to the full model. For example, on the Thoracic dataset, the mBO-V score was $32.9\%$, significantly lower than the $37.7\%$ achieved by the full model. This indicates that random feature masking is less effective than the structured masking strategy employed by the slot-BERT module.
}

\section{{Additional Results on Non-Surgical Data}}
\label{supple_natural}
{
We additionally evaluate Slot-BERT on non-surgical datasets, including the natural real-world dataset YT-VIS \mbox{\citep{xu2019youtubevis}} and the synthetic dataset MOVi-E \mbox{\citep{greff2022kubric}}. Following prior work \mbox{\citep{wu2023slotdiffusion,zadaianchuk2024object}}, we use $7$ slots for training on YT-VIS and $15$ slots for MOVi-E. For evaluation, we sample $6$ frames on YT-VIS and $24$ frames on MOVi-E, consistent with the setup in Video-Saur \mbox{\citep{zadaianchuk2024object}}. 

As shown in Table \ref{tab_ytvis_mov}, on the real-world YT-VIS dataset our method outperforms the second-best Video-Saur by over $+6\%$ mBO-V. On the synthetic MOVi-E dataset, Slot-BERT achieves comparable FG-ARI performance to Video-Saur. We note that MOVi-E provides less challenging conditions for object discovery due to its fixed camera viewpoint and relatively simple backgrounds which likely reduces the performance gap between methods.
}
\begin{table}[t!]
\caption{{Additional results on non-surgical datasets YT-VIS~\mbox{\citep{xu2019youtubevis}} and MOVi-E~\mbox{\citep{greff2022kubric}}. Results are reported in terms of mBO-V and FG-ARI. Best results are highlighted in bold.}}
\label{tab_ytvis_mov}
\centering
\setlength\arrayrulewidth{0.9pt}
\setlength\doublerulesep{0.9pt} 
\resizebox{1.0\linewidth}{!}{%
\begin{tabular}{lcccc}
\hline
\multirow{2}{*}{Method} & \multicolumn{2}{c}{YT-VIS} & \multicolumn{2}{c}{MOVi-E} \\ \cline{2-5}
 & mBO-V & FG-ARI  & mBO-V  & FG-ARI \\ \hline
SAVi~\citep{kipf2021conditional}           & 13.6 & 22.2 & 23.5 & 46.7 \\ 
STEVE~\citep{singh2022simple}              & 26.5 & 36.1 & 27.9 & 56.0 \\ 
Slot-Diffusion~\citep{wu2023slotdiffusion} & 28.9 & 14.7 & 30.2 & 60.0 \\ 
Video-Saur~\citep{zadaianchuk2024object}   & 29.1 & 39.5 & \textbf{35.6} & 73.9 \\ 
Slot-BERT (ours)                           & \textbf{35.3} & \textbf{41.1} & 35.4 & \textbf{74.3} \\ \hline
\end{tabular}
}
\end{table}

\section{Additional qualitative results}

\subsection{Comparison on using MLP and slot-Mixer decoder}
\label{sec_quali_mlp_mix}

In Section \ref{result_mixer}, we present the results of using the alternative Slot-Mixer decoder instead of the MLP decoder (as explained in Section \ref{method_decoder}) to reconstruct target features, exploring various setup variations. Detailed qualitative comparisons between different decoders are provided here. Figure \ref{fig_miccaisupp} illustrates segmentation masks produced by Slot-BERT with both MLP and Slot-Mixer decoders, tested on sequences of 30 frames at 1 FPS from the MICCAI dataset. The model was trained using 7 slots.

Both MLP and Slot-Mixer decoders demonstrate unique strengths. With the MLP decoder, the segmentation masks tend to cover objects more comprehensively and accurately locate them. On the other hand, the Slot-Mixer decoder exhibits superior temporal consistency {, since its within-frame attention at decoding provides synergistic benefits when combined with the temporal attention (TST), further stabilizing segmentation across video frames}. For instance, in the first and third rows of Figure \ref{fig_miccaisupp}, the instrument undergoes significant motion (e.g., moving across the image within a few frames). While the MLP decoder loses track of the object during such maneuvers, the Slot-Mixer decoder successfully maintains object tracking. This is evident in the image sequence, where the Slot-Mixer preserves the correct mask order despite the large motion, showcasing its robustness in maintaining temporal coherence.

\subsection{Zero-shot Segmentation of longer video sequence}

As discussed in Section \ref{result_0_transfer} of the main text, the model trained on MICCAI abdominal data demonstrates strong performance on thoracic data, which involves similar instruments but different tissue backgrounds. Figure \ref{fig_0thora} presents qualitative segmentation results for the zero-shot transfer of Slot-BERT with the Mixer decoder applied to thoracic robotic surgery videos.

Despite the differences in tissue backgrounds, the model effectively segments and tracks instruments. Notably, a slot successfully attends to gauze, an object not present in the MICCAI dataset, as seen in the first, second, and sixth rows. For instruments that are common between the MICCAI abdominal surgery data and the thoracic videos, such as needle drivers, the model achieves superior segmentation results, showcasing its ability to generalize across different surgical contexts.




\end{document}